\documentclass[reprint,amsmath,amssymb,superscriptaddress,longbibliography,aps,pra]{revtex4-1}
\usepackage{graphicx}
\usepackage{dcolumn}
\usepackage{subfigure}
\usepackage{color}
\usepackage{soul}
\usepackage[pdftex,breaklinks,colorlinks,
linkcolor=blue,
anchorcolor=blue,
citecolor=blue,
urlcolor=blue,]{hyperref}
\usepackage[bottom]{footmisc}
\begin{document}


\title{Methane storage in nanoporous media as observed via high field NMR relaxometry}

\author{A. Papaioannou}
\affiliation{Schlumberger Doll Research, Cambridge, MA 02139, USA}
\affiliation{City University of New York, The Graduate Center, Department of Physics, New York, NY 10016, USA}

\author{R. Kausik}
\email{RViswanathan@slb.com}
\affiliation{Schlumberger Doll Research, Cambridge, MA 02139, USA}

\date{\today}
\begin{abstract}

The storage properties of methane gas in Vycor porous glass (5.7 nm) are characterized in a wide pressure range from 0.7 MPa-89.7 MPa using Nuclear Magnetic Resonance (NMR). We demonstrate the capability of high field NMR relaxometry for the determination of the methane gas storage capacity and the measurement of the Hydrogen Index, to a high degree of accuracy. This helps determine the excess gas in the pore space which can be identified to exhibit Langmuir properties in the low pressure regime of 0.7 MPa to 39.6 Mpa. The Langmuir model enables us to determine the equilibrium density of the monolayer of adsorbed gas to be 8.5\% lower than that of liquid methane. We also identify the signatures of multilayer adsorption at the high pressure regime from 39.6 Mpa to 89.7 Mpa and use the Brunauer-Emmet-Teller (BET) theory to determine the number of adsorbed layers of methane gas. We show how these measurements help us differentiate the gas stored in the Vycor pore space into free and adsorbed fractions for the entire pressure range paving way for similar applications such as studying natural gas storage in gas shale rock or hydrogen storage in carbon nanotubes.

\end{abstract}

\maketitle

\section{Introduction}\label{sec:intro}
	
	Nuclear magnetic resonance (NMR) is a well-established technique for probing the dynamics of fluids in porous media with disordered geometries, such as rocks and biological systems \cite{mair1999probing,boutis2007high}. NMR investigations of fluids in porous media are generally performed through measurements of relaxation times and diffusion coefficients. This enables the determination of the fluid properties such as viscosity, chemical composition and fluid-pore surface interactions. In addition, the geometrical properties of the porous matrix such as tortuosity ($\alpha$), surface to volume ratio ($S/V$) and diffusive permeability may also be determined by studying the relaxation and time-dependent diffusion coefficient of the confined fluid \cite{novikov2012characterizing,mitra1992diffusion,diffraction}. For these reasons, NMR has become an invaluable tool, especially in oil and gas exploration applications  where it is routinely applied for downhole logging and laboratory rock core analysis. NMR has also found  broader applications in energy research including investigations of methane gas in shale rocks \cite{wang2014high,gasparik2012high,kausik2011characterization} and hydrogen storage in carbon nanotubes and other nanoporous materials \cite{panella2005hydrogen,horch2014high}.
	
	The storage of gas molecules in porous media has seen renewed interest especially due to recent advances in natural gas production from gas shale reservoirs. Gas shales are organic-rich mudstones characterized by the presence of nanometer size organic kerogen pores, hosting natural gas. The gas molecules in the nanopores coexist in two phases: free gas that has bulk-like properties and adsorbed gas that undergoes strong interactions with the pore walls \cite{kausik2011characterization}. These two phases are in fast exchange during the time scales of the NMR experiment \cite{bear2013dynamics}. In addition, the nature and dynamics of these two phases depend on the temperature and pressure of the system, making adsorption studies of great importance for understanding natural gas storage and extraction. Though NMR of fluids in porous media is a well-established field covering studies such as those of diffusion of supercritical fluids in nanoporous media using pulsed field gradient techniques \cite{dvoyashkin2007supercritical,valiullin2011impact,valiullin2009correlating} and methane gas adsorption on graphite \cite{riehl1972nmr}, the storage properties of natural gas in nano pores at pressures close to the highest pressures encountered in gas shale reservoirs ($\sim90$ MPa) have not been studied to our knowledge.
	
In this work we describe a detailed study of methane gas storage in nanoporous Vycor glass. We confirm spin-rotation to be the dominant contributor to the relaxation Hamiltonian of bulk methane gas, up to ultra-high pressures of $70$ MPa. We demonstrate the capability of high field NMR to determine the gas capacity, Hydrogen-Index (HI) (sec. \ref{sec:Hydrogen_Index}), adsorbed gas density, and enable the separation of free and adsorbed gas quantities in Vycor nanoporous glass at a wide range of pressures ($0.7$ MPa-$89.6$ MPa). We show how this analysis is enabled by the unique spin-spin relaxation times of the methane gas in the pore space. The quantification of methane gas stored in the nanopores together with the pore volume allow for a direct measurement of the effective Hydrogen-Index. We determine the excess gas in the pore space due to adsorption by using either monolayer or multilayer adsorption models for the appropriate low pressure and high pressure regimes. This enables the determination of the equilibrium density of the adsorbed layer and the number of adsorbed layers permitting the separation of the total gas into free and adsorbed fractions.
		

\section{Experimental Methods}\label{sec:Methods}

\begin{figure}
 	 \centering
  	\includegraphics[width=1.0\linewidth]{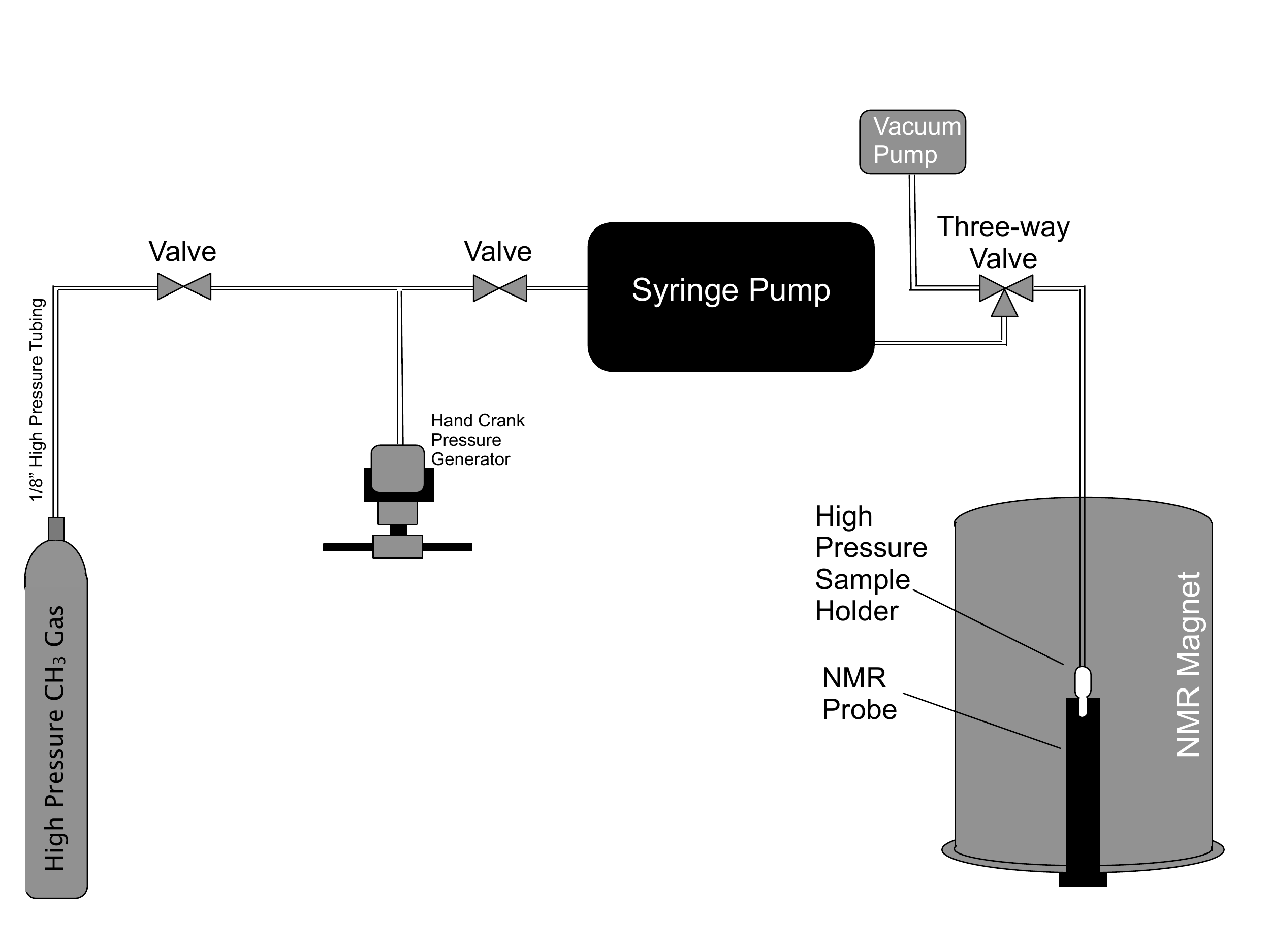}
 	 \caption{ Schematic representation of the experimental setup used in this work. The apparatus is composed of two pressure generators, the hand crank pump and the syringe pump. The upper pressure limit for this setup is $206.8$ MPa. The pressure is regulated digitally to within $0.35$ MPa.   }
	 \label{fig:Exp_setup} 				 
\end{figure}

All experiments were performed on a Bruker 400MHz system equipped with high-pressure apparatus as shown in the schematic in Fig. \ref{fig:Exp_setup}. The main pressure generation was carried out through the syringe pump and the hand crank pressure generator, which were connected to the high pressure NMR sample tube. The hand crank pressure generator had a volume of 18 ml and was purchased from High Pressure Equipment Company (HIP). The syringe pump (Daedalus Innovations LLC) with a capacity of 6.7 ml and pressure rating of $413.7$ MPa was used for secondary compression. The NMR sample tube (Daedalus Innovations) made of zirconium, had outer and inner diameters of 5.0 mm and 3.0 mm respectively, and was 92 mm long. Methane gas of 99.9\% purity was purchased from Air-Gas (Radnor Township-PA) in a cylinder pressurized to $13.8$ MPa. A three way valve was used to direct the methane gas to either the sample tube or the vacuum pump. All the connections were well evacuated before the experiments to purge out all oxygen in the system and avoid any additional relaxation effects. The high pressure components were connected with pressure tubing and valves rated to $413.7$ MPa. 

Vycor nanoporous glass manufactured by Corning Inc (New York, USA),  was purchased in the form of cylindrical rods of diameter 2.96 mm. They were cut into samples of 11.50 mm to fit in the center of the NMR coil (16.68 mm). The pore size of the Vycor rods was 5.7 nm and the porosity 28\% of the sample volume, as determined by BJH adsorption studies.

\begin{figure}
 	 \centering
  	\includegraphics[width=1.0\linewidth]{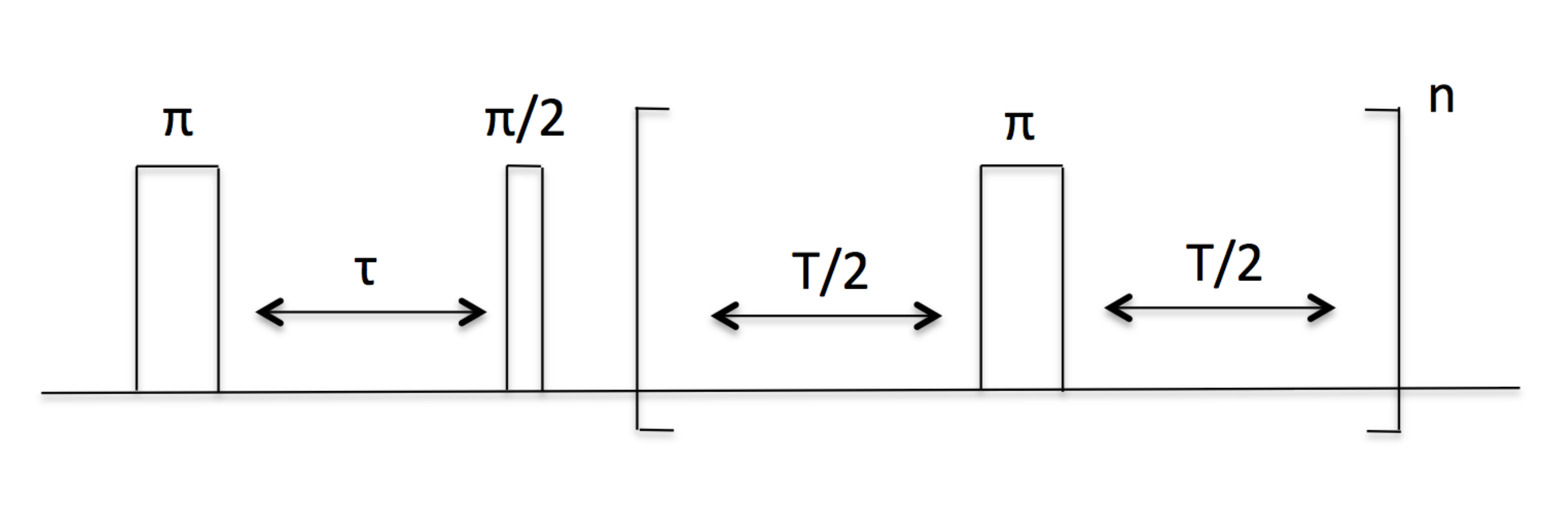}
 	 \caption{ Two dimensional $T_1-T_2$ correlation pulse sequence used in this work. As discussed in the text, the time interval $\tau$ in the indirect dimension was varied from 10ms to 60s. The number of $\pi$ pulses was 32,000 and the time echo time T was 200$\mu s$.    }
	 \label{fig:ppg} 				 
\end{figure}

The transverse relaxation times ($T_2$) were measured using the Carr-Purcell-Meiboom-Gill (CPMG) pulse sequence \cite{carr1954effects,meiboom1958modified}. The two dimensional $T_1$-$T_2$ correlation experiments were performed by using the inversion recovery pulse sequence in the indirect dimension and the CPMG pulse sequence in the direct dimension, as shown in Fig. \ref{fig:ppg}. The time interval $\tau$ for encoding in the indirect dimension was varied from 10ms to 60s (the time at which the magnetization is fully relaxed). The number of $\pi$ pulses was chosen to be 32,000 to cover the entire $T_2$ spectrum and the echo time T was 200$\mu s$. Diffusion effects were confirmed to be negligible at such short echo times by observing the attenuation of the first five echoes at various echo times \cite{hurlimann1994restricted}. The last delay (repetition time) between scans was varied for each pressure from 20s to 60s. The temperature was kept constant at 22$^{\circ}$C throughout all the experiments with an accuracy of $\pm1^{\circ}$C. 

The RF probe and the receiver were calibrated to measure the number of $^{1}H$ nuclei in the sample by carrying out $T_2$ measurements of 5$\mu L$ to 40$\mu L$ of $H_2O$ samples at 22$^{\circ}$C and atmospheric pressure. We determined the accuracy of the calibration to be within 4\%  by comparing the NMR signals of bulk methane gas at various pressures with its density values from National Institute of Standards and Technology (NIST)\cite{linstrom2001nist}. 


\section{Results and discussion}\label{sec:results}
	\subsection{Relaxation of free methane gas}
	\label{Sec:T1_T2_Bulk}
\begin{figure}
 	 \centering
  	\includegraphics[width=1.1\linewidth]{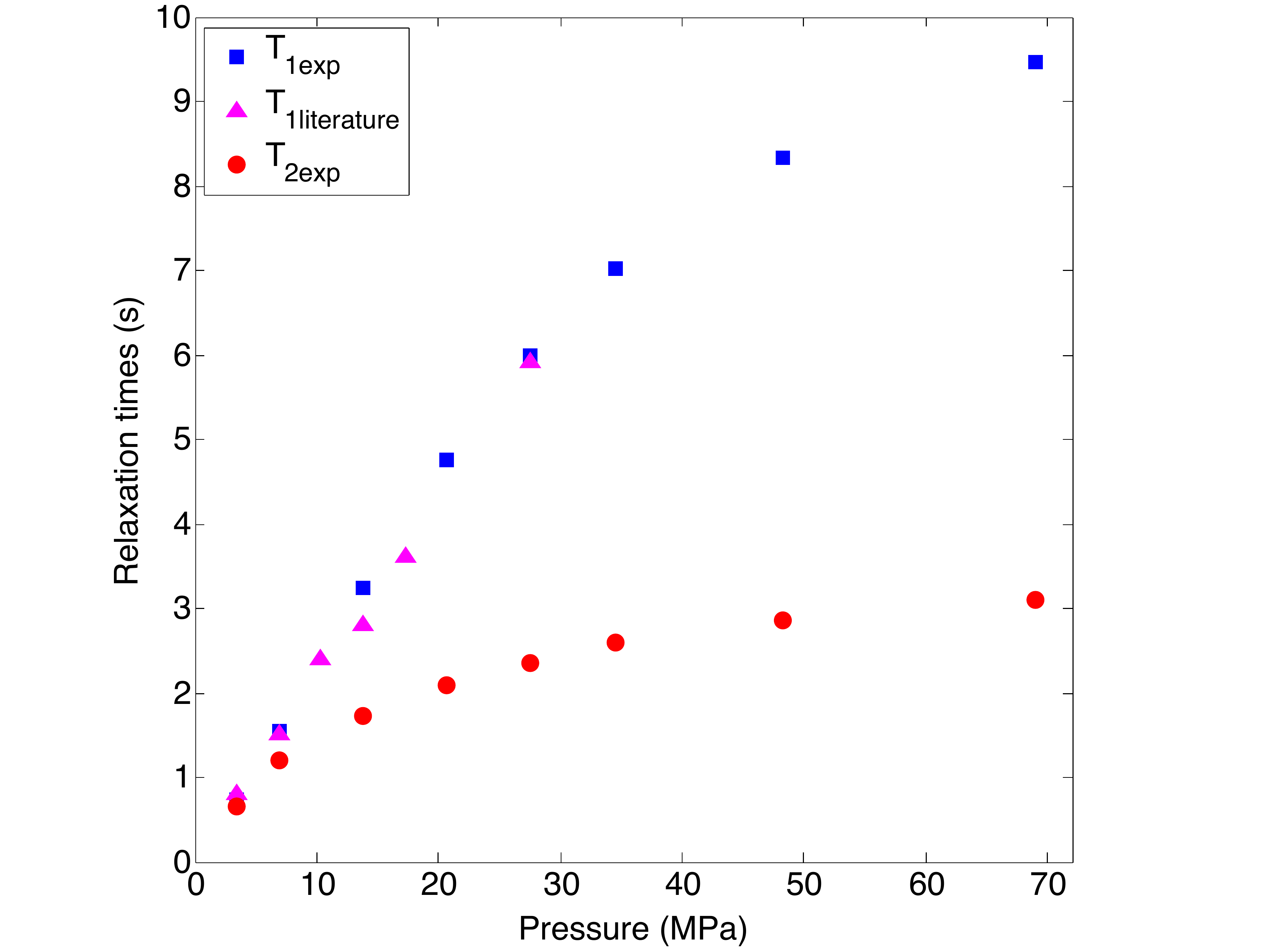}
 	 \caption{Longitudinal ($T_1$) and transverse ($T_2$) relaxation times of bulk methane gas at various pressures at 22$^{\circ}C$. To convert the magnetization decays into relaxation distributions, a fast inverse Laplace transform algorithm was used \cite{venkataramanan2002solving} and each experimental point corresponds to the log mean of the entire 1D distribution. The reduced value of $T_2$ in comparison to $T_1$ is because of the interactions with the sample tube walls. The literature values of $T_{1literature}$ (magenta) are taken from Ref. \cite{hari1998nmr}.   }
	 \label{fig:T1T2_freemethane} 				 
\end{figure}

	The transverse ($T_2$) and longitudinal ($T_1$) relaxation times of bulk methane gas were measured for a wide range of pressures from $3.5$ MPa to $68.9$ MPa at 22$^{\circ}$C. The critical pressure and temperature of methane are $4.6$ MPa and -$82.75^{\circ}$C respectively, and therefore most experiments were conducted in the super critical state. The measured $T_1$ relaxation times are consistent with the literature data at the low pressure range measured by Watson et al. \cite{hari1998nmr} as shown in Fig. \ref{fig:T1T2_freemethane}. The relaxation times of bulk methane gas are dominated by the spin rotation mechanism at low to moderate pressures while at high pressures, due to the increased density, the mean free path decreases and inter molecular dipolar interactions could start playing a role \cite{johnson1961nuclear}.

In the motional narrowing limit the relationship between the relaxation rates is given by,
\begin{equation}
	T_1^{-1}=T_2^{-1}=\frac{2(C^2_{\parallel}+2C^2_{\perp})I_1\tau_F}{3\hbar^2}kT.
	\label{eq:relaxrates}
\end{equation}

In the above expression $\tau_F$ is the rotational correlation time, $k$ is the Boltzmann's constant, $I_1$ is the moment of inertia of the spherical molecule, $T$ is the temperature and $C_{\parallel}$, $C_{\perp}$ are the principal components of the spin rotation tensor \cite{hubbard1963theory}. The correlation time, $\tau_F$ is given by,
\begin{equation}
	\tau_F=\frac{3I_1D}{4a^2kT},
\end{equation}
where $D$ denotes the diffusion coefficient of the molecule which depends on the viscosity as $D=kT/6\pi a \eta$, with $a$ being the radius of the molecule and $\eta$ the viscosity.

Bulk methane gas is in the motional narrowing limit at the Larmor frequency of the experiments ($f = 400MHz$) and therefore $T_1$ should be equal to $T_2$. However, as shown in Fig. \ref{fig:T1T2_freemethane} the $T_2$ values are consistently shorter than the $T_1$ values at all measured pressures. This is because the spin-spin relaxation times ($T_2$) are sensitive to the slow motions of methane molecules caused by their interaction with the NMR sample tube walls. This may be understood by inspecting the frequency dependence of the relaxation rates \cite{abragam1961principles},
 
 \begin{equation}
 \begin{split}
 	T_1^{-1}(\omega) =\left( \frac{\mu_0}{4\pi}\right)^2 \frac{3\gamma^4\hbar^2I(I+1)}{2r^6}[\mathcal{J}^{(1)}(\omega)\\
	+\mathcal{J}^{(2)}(2\omega)],
	\end{split}
\label{eq:R1}
 \end{equation}
and,
 \begin{equation}
 \begin{split}
 	T_2^{-1}(\omega)  =\left( \frac{\mu_0}{4\pi}\right)^2 \frac{3\gamma^4\hbar^2I(I+1)}{2r^6}[\frac{1}{4}\mathcal{J}^{(0)}(0)\\
	+\frac{5}{2}\mathcal{J}^{(1)}(\omega)+\frac{1}{4}\mathcal{J}^{(2)}(2\omega)],
	\end{split}
\label{eq:R2}
 \end{equation}

where $\mu_0$ is the vacuum permeability, $I$ is the spin number, $\gamma$ is the gyromagnetic ratio and $r$ is the internuclear vector between the $^{1}H$ nuclei.
The spectral densities $\mathcal{J}^{(n)}(\omega)$ can be obtained by the Fourier transform of the autocorrelation function $\mathcal{G}(t)\equiv \langle B(t)B(t+\tau)\rangle$ which describes the time dependent fluctuations of the local magnetic field $B(t)$. The $T_2$ relaxation times are dominated by the $\mathcal{J}^{(0)}(\omega=0)$ term and are therefore very sensitive to the low frequency or slow motions while the $T_1$ relaxation times are sensitive to the much higher Larmor frequency ($\omega_0$ \& $2\omega_0$).

Assuming Gaussian diffusion in 1D, the mean displacement of the methane gas molecules may be calculated as,
$\langle\Delta x(t)\rangle = (2Dt)^{1/2}$, where D is the bulk diffusion coefficient of methane and t is the diffusion time period. At a pressure of $20.7$ MPa the spins have a self-diffusion coefficient of $7.7\cdot10^{-8}m^2/s$ \cite{oosting1971proton} and a spin lattice relaxation time of $5.2s$ \cite{gerritsma1971proton}, resulting in displacements of 0.9 mm during their $T_1$ lifetimes. As the high pressure tube has an inner diameter of 3.0 mm, a fraction of the spins interact with the tube walls. This makes the magnetization decay of the $T_2$ measurement (CPMG sequence) multi-exponential. The $T_2$ relaxation times shown in Fig. \ref{fig:T1T2_freemethane} are the log mean values of the $T_2$ distributions.  The spin lattice relaxation times ($T_1$), being only sensitive to motions that occur around the Larmor frequency ($f = 400MHz$), are not affected and the corresponding magnetization decays are mono exponential.

\begin{figure}
 	 \centering
  	\includegraphics[width=1.1\linewidth]{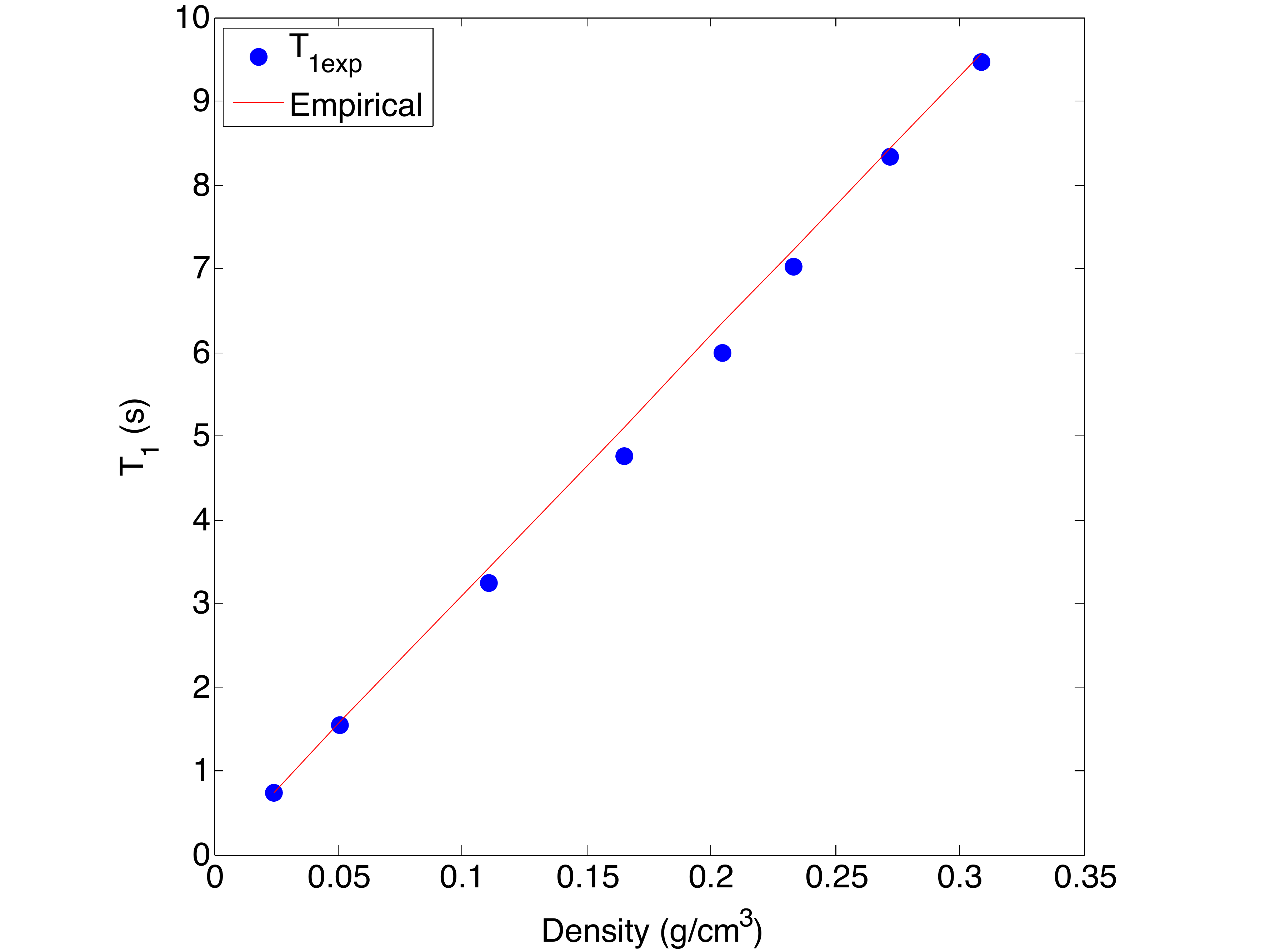}
 	 \caption{Longitudinal ($T_1$) relaxation times of bulk methane gas at various densities at 22$^{\circ}C$. The 1D data were processed using the ILT algorithm \cite{song2002t} and each experimental point (blue circles) corresponds to the log mean of the entire 1D distribution. The solid red line corresponds to the empirical prediction of $T_1$ relaxation times for spin rotation interaction taken from \cite{sho2000correlations}. As discussed in the text, the empirical prediction fits the data implying that spin rotation interactions dominate the relaxation Hamiltonian at even such high densities for bulk methane.}
	 \label{fig:T1T2_hirasaki} 				 
\end{figure} 

The total Hamiltonian for the spin ensemble in the rotating frame may be written as, $\mathcal{H}_{total} = \mathcal{H}_{ij}^{DD} + \mathcal{H}_{SR}$ where $\mathcal{H}_{SR}$ describes the spin rotation interaction and $\mathcal{H}_{ij}^{DD}$ is the dipolar Hamiltonian respectively. At the pressure of $68.9$ MPa the density of bulk methane is still 26\% lower than that of liquid methane. At low pressures where the total Hamiltonian is completely dominated by spin rotation interaction an empirical relation between the spin lattice relaxation times and the methane gas density was obtained \cite{sho2000correlations},

\begin{equation}
	T_{1SR}=c\cdot\rho\cdot T^{-3/2},
	\label{eq:Hirasaki}
\end{equation}

where $\rho$ is the density in $g/cm^3$, $T$ is the temperature in $K$ and the constant $c=1.57\cdot10^{5}[cm^3K^{3/2}sg^{-1}]$. This empirical relation was obtained by Sho-Wei et al. \cite{sho2000correlations} using experimental data of longitudinal relaxation times of methane gas for pressures up to approximately 20 MPa.

In Fig. \ref{fig:T1T2_hirasaki} we compare our measured $T_1$ relaxation times with  the computed values using the empirical relation given by eq. \ref{eq:Hirasaki} for the entire pressure range from $3.4$ MPa to $68.9$ MPa. Excellent agreement ($\chi^2/dof\simeq 1$ \cite{bevington1969data}) was found between the empirical prediction of eq. \ref{eq:Hirasaki} right from lower pressures up to $68.9$ Mpa (0.307$g/cm^3$) demonstrating that $\mathcal{H}_{total}$ is dominated  by the spin rotation interaction throughout this pressure range. It should be noted here that no phase transition effects were observed when traversing the critical pressure ($P_c=4.6$ MPa) at $T=22^{\circ}C$ as indicated by the $T_1$ and $T_2$ relaxation times. Previous studies across the critical point (traversing both the critical pressure and temperature) showed changes in the relaxation times due to critical density fluctuations \cite{krynicki1977proton, topol1979nuclear} and distribution of correlation times \cite{resing1965apparent}, but such effects were not seen as we are far from the critical points in our experiments.


\subsection{Relaxation of methane gas in Vycor glass}

Vycor porous glass is composed of of 96\% $SiO_2$ and 4\% $Br_2O_3$ and the samples used in this study had a porosity of 28\% and an average pore size of 5.7 nm. This results in a high surface to volume ratio of $1.25nm^{-1}$ that leads to enhanced surface interaction of the confined methane molecules with the pore walls. The NMR relaxation of spin 1/2 nuclei in porous media generally occurs through two dominant pathways. The first is the slow motions at the surfaces resulting in enhanced relaxation caused by mechanisms such as Reorientation Mediated by Translational Diffusion (RMTD) \cite{kimmich1993nmr}. This effect is due to the geometrical characteristics of the confining system. The second is the dipolar relaxation due to the presence of paramagnetic impurities in the porous medium. As Vycor does not host significant quantities of paramagnetic impurities \cite{levitz2000probing}, the methane gas relaxation in this case is expected to be dominated by the slow motions at the pore walls.


\subsubsection{2D $T_1-T_2$ relaxation}
\label{sec:T1_T2}

\begin{figure}
 	 \centering
  	\includegraphics[width=1.1\linewidth]{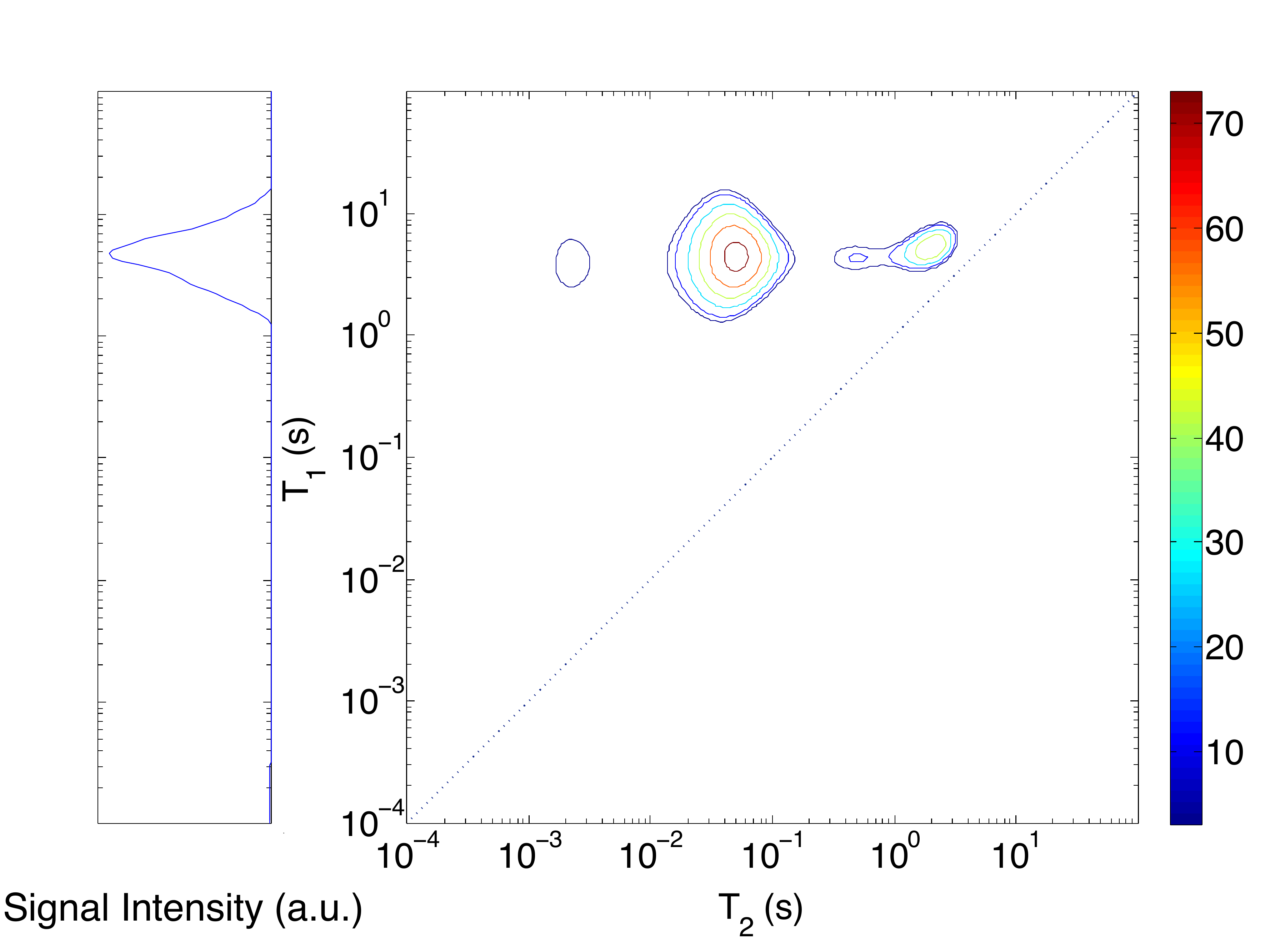}
 	 \caption{$T_1-T_2$ correlation map for methane gas in Vycor glass at a pressure of $27.6$ MPa and temperature of 22$^{\circ}C$ is shown. While only one peak can be resolved in the $T_1$ dimension, the $T_2$ dimension allows for separation of different components in the form of three different peaks. The right most peak correspond to the annulus gas while the center and the left most correspond to the methane gas in the Vycor pore space. The intensity bar on the right is in [a.u].}
	 \label{fig:2DT1T2_freemethane} 				 
\end{figure} 

The results of the $T_1-T_2$ correlation experiments of methane in Vycor at $27.6$ MPa are shown in Fig. \ref{fig:2DT1T2_freemethane}. The relaxation times were measured using an inversion recovery pulse sequence followed by the CPMG sequence as shown in Fig. \ref{fig:ppg} \cite{callaghan2011translational} and the results were analyzed using an Inverse Laplace Transform (ILT) algorithm \cite{venkataramanan2002solving}. Three well distinguishable peaks were observed in the direct dimension ($T_2$). On the other hand, the indirect dimension shows a single $T_1$ peak at a value of about 4.0s. This is because the $T_1$ relaxation mechanism is only sensitive to motions at the Larmor frequency ($f = 400MHz$) and is therefore not affected by the slow motions of methane gas in the different environments. 	 

The different peaks in the $T_2$ dimension are a result of the methane molecules in different environments experiencing different relaxation pathways. In the following sections we identify the origin of the different components of relaxation in Vycor glass and use this information to quantitatively assess the adsorption processes at the high and low pressure regimes. 
  

\subsubsection{Spin-spin relaxation}
\label{sec:T2_pressure}

\begin{figure}
 	 \centering
  	\includegraphics[width=1.1\linewidth]{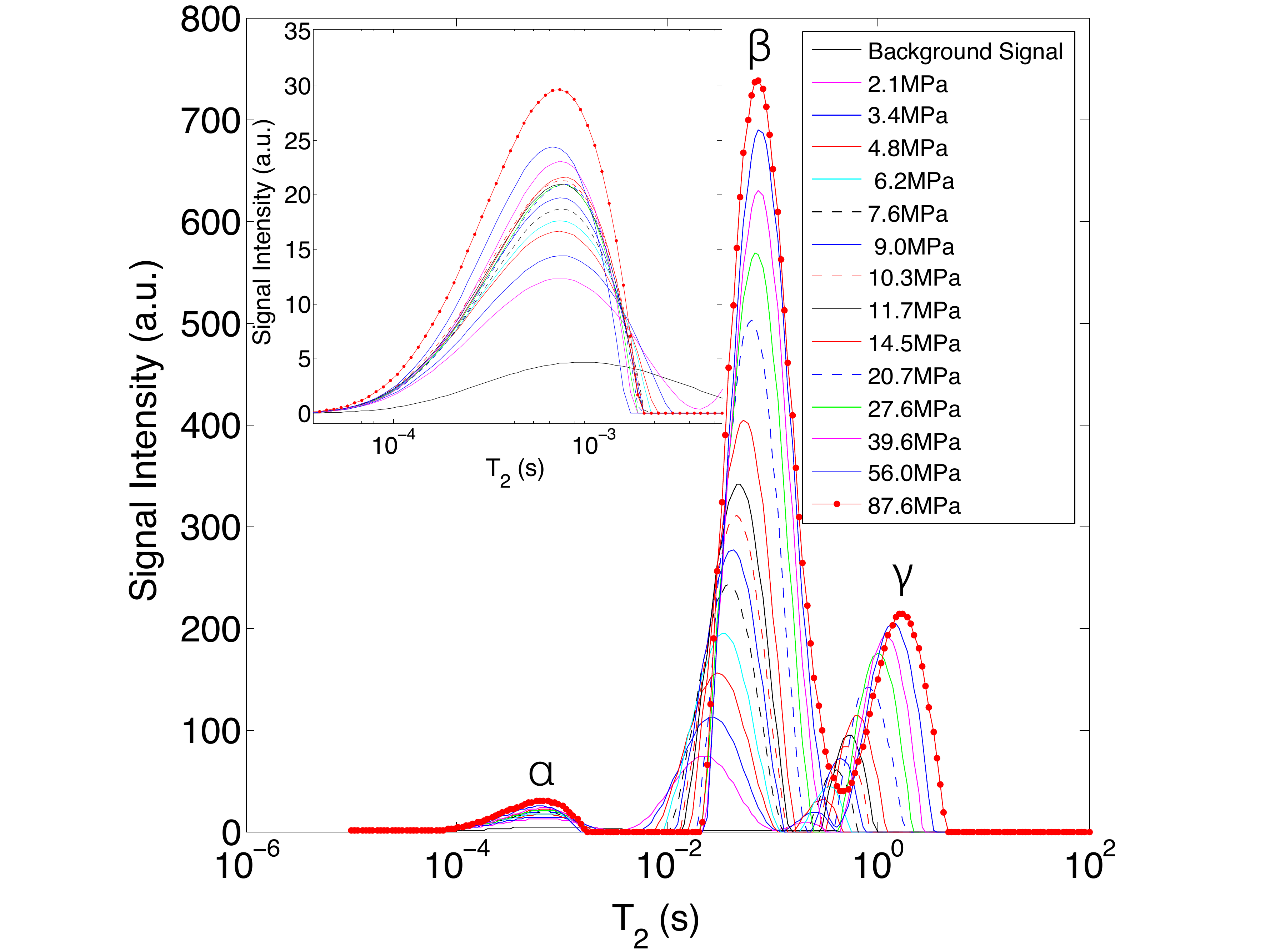}
 	 \caption{Transverse relaxation times ($T_2$) of methane gas in Vycor are shown for various pressures at a constant temperature of 22$^{\circ}C$. The pressure dependence of the signal intensity and the $T_2$ values for the different components can be identified. The $\alpha$ component is zoomed in the inset to highlight the background signal of Vycor.} 
	 \label{fig:T2_distributions} 				 
\end{figure}

The pressure dependence of the $T_2$ relaxation times of methane gas in Vycor is shown in Fig. \ref{fig:T2_distributions}. The peak with the longest relaxation time, labeled $\gamma$ in Fig. \ref{fig:T2_distributions}, corresponds to the free methane gas in the annulus of the NMR tube (between the Vycor cylinder and the sample holder) and above the Vycor glass cylinder. This free methane gas relaxes mainly via spin rotation interaction resulting in relaxation times comparable to those of the bulk gas. The small difference in the relaxation times of these methane molecules from that of the bulk ($2.4s$ instead of $2.9s$ at $27.6$ MPa) is due to surface relaxation occurring on the Vycor rod surface and the NMR tube walls. The identity of this phase was further confirmed by carrying out experiments with a teflon tape covering the annulus space, resulting in drastic reduction of this long relaxing peak.

The $\beta$ component of Fig. \ref{fig:T2_distributions} corresponds to the methane gas in the pore space of Vycor. The significant reduction in the spin-spin relaxation time of the $\beta$ peak with respect to the annulus gas ($\gamma$ component) is because of the enhanced surface interaction of the methane molecules with the pore walls. The strong pressure dependent increase in the intensity of this peak compared to bulk methane gas is due to adsorption of methane molecules on the pore walls. Adsorbed gas has much higher density compared to bulk gas resulting in higher spin density in the pores. As the time scale of exchange between the free and adsorbed gas is much faster than the NMR relaxation time scale \cite{bear2013dynamics}, these two components cannot be separated and result in a single $T_2$ distribution for the free and adsorbed phases. The $\alpha$ component corresponds to the small number of methane gas molecules trapped in pore spaces of smaller dimensions. Vycor porous glass has a pore size distribution that peaks at 5.7 nm and also hosts a few smaller pores due to geometrical disorder. These molecules undergo even higher surface interactions resulting in relaxation times on the order of 1ms. The smaller tortuous, dead-end channels can also reduce molecular mobility \cite{dozier1986self}. A small background signal with the same relaxation time as the $\alpha$ component was observed before gas injection which may correspond to trapped water molecules that could not be evacuated by the vacuum procedure and is shown as an inset in Fig. \ref{fig:T2_distributions}.

\subsubsection{2D $T_2-T_2$ relaxation}

To further understand the dynamics of the different $T_2$ distributions, diffusive coupling of the methane gas between the different environments was probed through the two dimensional $T_2-T_2$ correlation measurements \cite{van2010analysis,lee1993two,washburn2006tracking}. This experiment consists of the measurement of spin-spin relaxation time in the indirect dimension (``evolution time") using a CPMG sequence, followed by storage of the magnetization in the z-axis (``storage time"), during which the spins relax by $T_1$ relaxation. The latter is followed by measurement of the spin-spin relaxation time during acquisition using another CPMG sequence. The number of CPMG echoes in the indirect dimension was varied to acquire a 2D $T_2-T_2$ plot for different storage times. As the spin lattice relaxation times ($T_1\sim 2s$) at $400MHz$ are very long compared to the $T_2$, the methane gas molecules could diffuse from one environment to another during the storage intervals resulting in non diagonal peaks in the 2D $T_2-T_2$ map. 

\begin{figure}
		\includegraphics[width=1.1\linewidth]{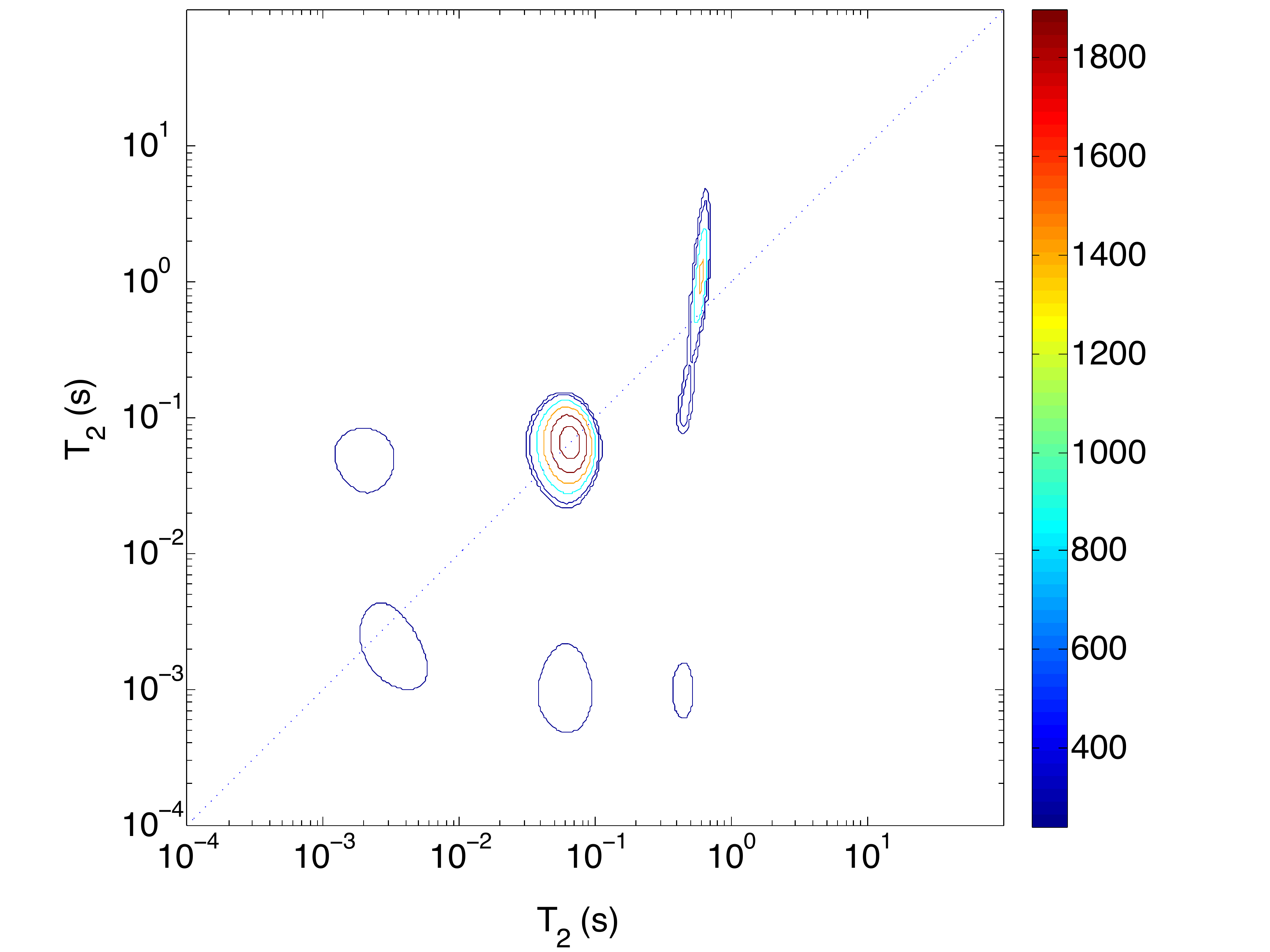}
   \caption{Two dimensional map of $T_2-T_2$ correlation experiment of methane in Vycor glass acquired with a storage time of 100ms. The diffusive coupling between different geometrical compartment results in non diagonal cross peaks. The first two peaks with the shorter relaxation times in the diagonal axis correspond to the gas stored in the Vycor pore space while the right most peak correspond to the gas in the annulus space.}
\label{fig:T2T2}
\end{figure}

The exchange between two coupled environments may be quantified via the variation of the cross peak intensities with respect to the storage time, only in the slow exchange limit, i.e. the exchange is slow during the evolution times \cite{van2010analysis}. This condition is not satisfied in our experiments with significant exchange occurring between the coupled environments during the CPMG sequence (evolution time), and therefore accurate exchange times could not be extracted. However, the experimental data shown in Fig. \ref{fig:T2T2} suggest that all three environments are coupled at the time scales of $100ms$, as cross peaks appear in the $T_2-T_2$ correlation plots. This indicates that the different gas species are coupled and that the $\alpha$ and $\beta$ peaks together correspond to the gas molecules in the Vycor pore space and therefore should be integrated to obtain the total stored gas.

\subsection{Methane gas Hydrogen-Index in porous media}
\label{sec:Hydrogen_Index}

\begin{figure}
 	 \centering
  	\includegraphics[width=1.1\linewidth]{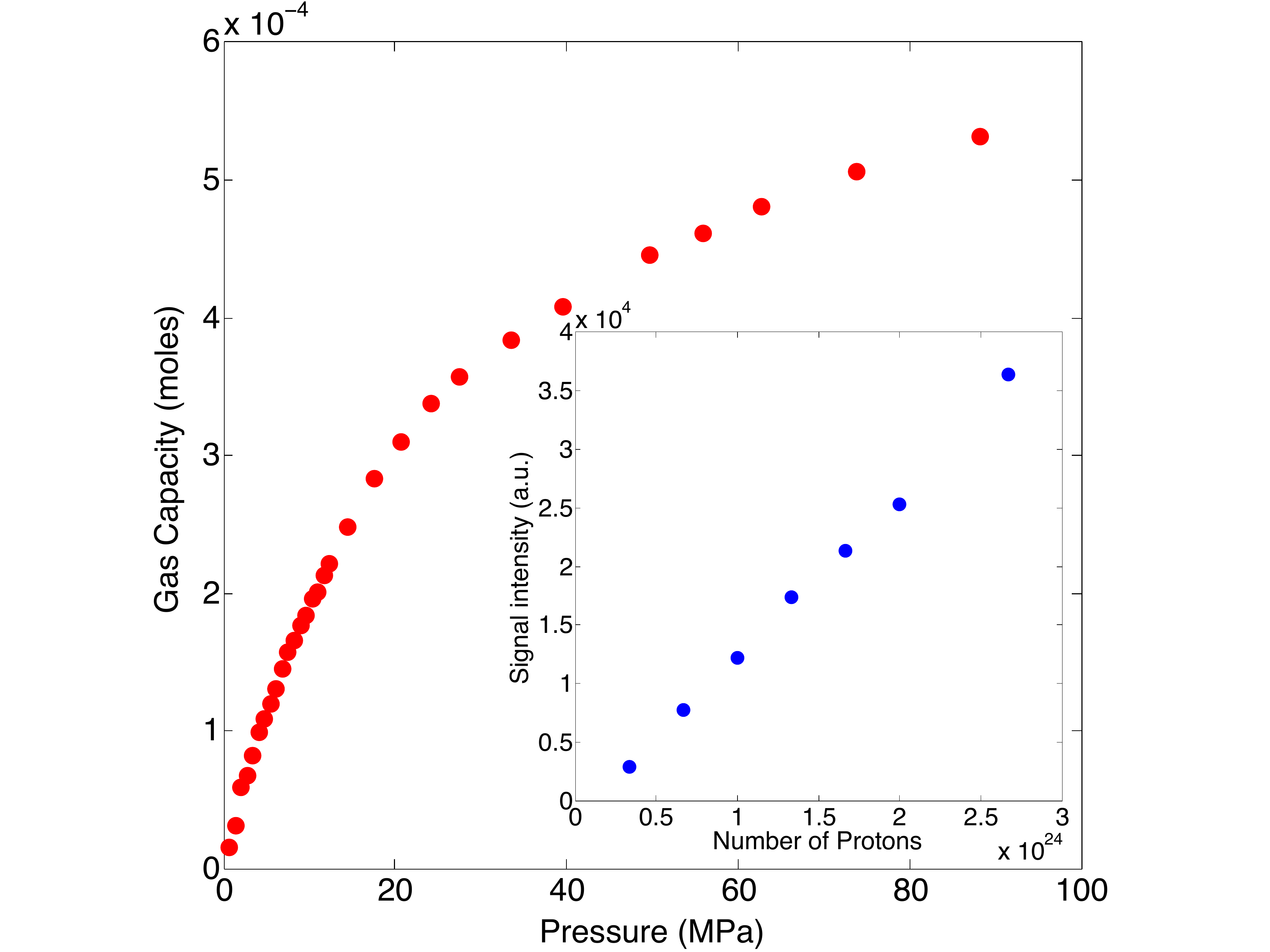}
 	 \caption{Methane gas stored in Vycor plotted as a function of pressure. The stored methane gas is shown to increase steeply at lower pressures and slowly saturates as the fluid density approaches that of liquid methane. The calibration of the signal intensity to the number of protons was carried out by using various volumes of H$_2$O and is shown in the inset.} 
	 \label{fig:GIP} 				 
\end{figure}

The gas stored in Vycor glass pore space is the sum of the $\alpha$ and $\beta$ $T_2$ distributions shown in Fig. \ref{fig:T2_distributions}. In Fig. \ref{fig:GIP} the methane gas stored in the Vycor pore space as a function of pressure is plotted. Note that the background signal as shown in the inset of Fig. \ref{fig:T2_distributions} was subtracted to obtain the signal from the gas stored in the Vycor pore space The stored gas was quantified using the calibration of the system to the signal from the $^{1}H$ nuclei in pure water. The results of the calibration of the system based on various volumes of pure water are shown as an inset of Fig. \ref{fig:GIP}. It should be noted that by using a Vycor sample 2.96 in diameter leads to a low filling factor. However, the results shown in Fig. \ref{fig:GIP} indicate that the low filling factor presents no challenge.

The Hydrogen Index (HI) of a fluid is an important parameter for oil and gas exploration as it is used for the interpretation of various downhole logging measurements, including NMR, to determine the quantity of fluids such as hydrocarbons and water. The HI is defined as the density of $^{1}H$ in the sample, relative to that of water \cite{kausik2011characterization}. In the case of gas filled rock samples the hydrogen index of the gas phase is given by \cite{kleinberg1996nmr},

\begin{equation}
	HI_g=\frac{\rho_g n_{Hg} }{\rho_w\cdot MW_g},
\end{equation}
where $\rho_w=$0.11 moles$\cdot$ cm$^{-3}$ and $\rho_g$ are the molar and mass densities of the water and gas, $MW_g$ is the molecular weight of methane gas and $n_{Hg}$ is the number of hydrogens in a single methane molecule ($n_{Hg}=4$).

\begin{figure}
 	 \centering
  	\includegraphics[width=1.1\linewidth]{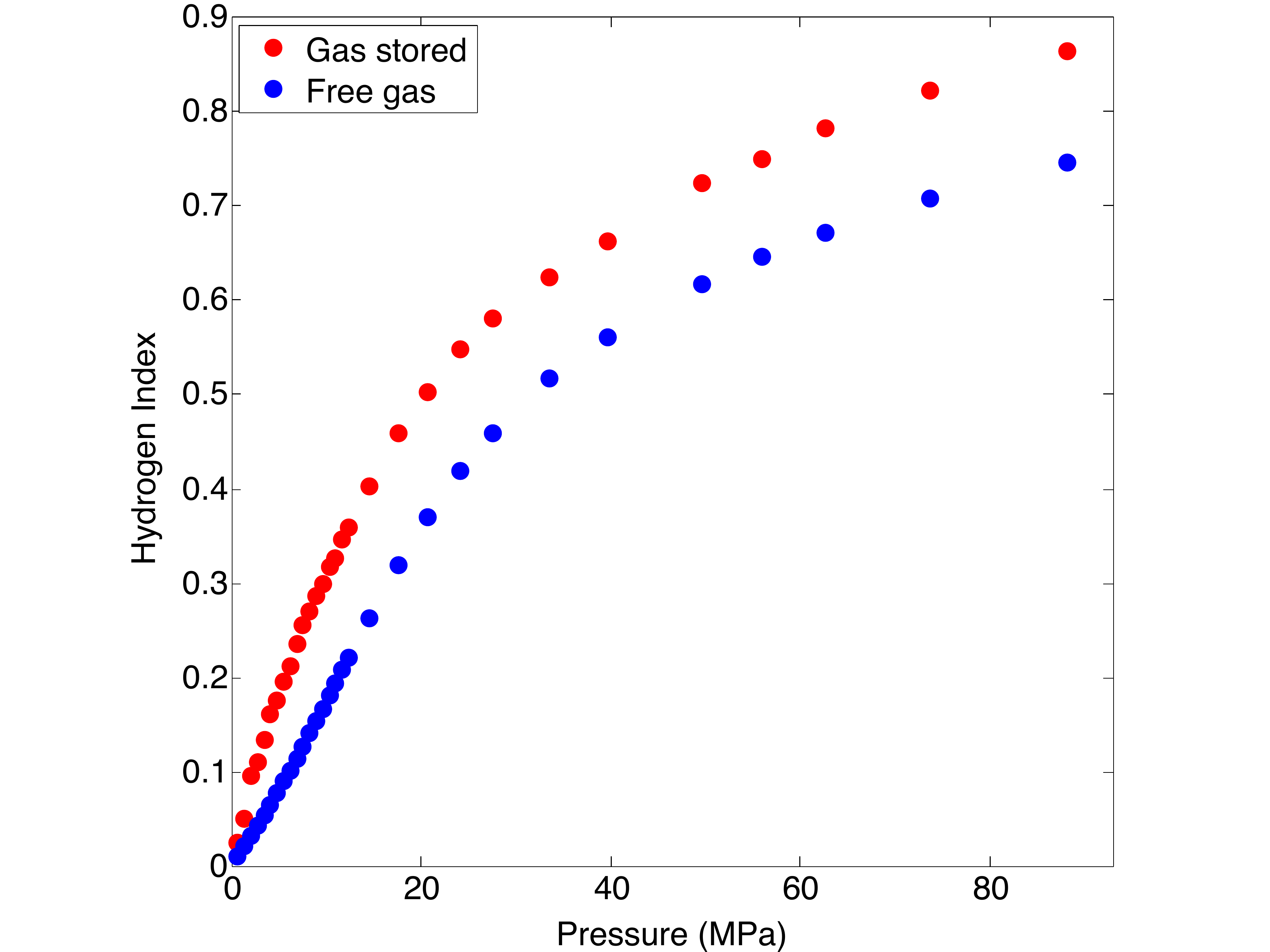}
 	 \caption{Effective Hydrogen index (HI) computed based on the measurement of the gas stored in the Vycor pore space. The free methane gas hydrogen index at various pressures is also shown as a comparison. } 
	 \label{fig:HI}				 
\end{figure}

While the HI and its pressure dependence for bulk oil, water and natural gas are well known \cite{kleinberg1996nmr}, the values for natural gas in shale nanopores are difficult to determine. This is because the gas in the shale pores exists as free and adsorbed phases whose densities also have a complex pressure dependence.

For petrophysical and log interpretation applications an effective HI is defined which is a weighted average of the free and adsorbed gas hydrogen indices in the porous medium. The hydrogen index of bulk methane gas occupying a pore volume equivalent to that of the Vycor sample (length of 11.50 mm, volume of 0.022 ml) is compared with the measured HI of methane gas in Vycor in Fig. \ref{fig:HI}. The effective Hydrogen Index of methane in the porous media is consistently higher than the bulk methane gas under the same conditions (pressure-volume-temperature) due to the increased density of the adsorbed gas. The measurement of the effective HI on real rock sample cores at formation temperatures and pressures using this methodology potentially enables direct log interpretation for obtaining fluid quantities.


\subsection{Monolayer and multilayer adsorption}

The gas inside the Vycor pore space exists in two phases, namely as adsorbed gas on the pore walls and as free gas in the center of the pore. As the free and adsorbed gas are in fast exchange in the NMR relaxation time scales (1$ms$ to 50$ms$), the measured $T_2$ relaxation times of the gas in the pores are a weighted average of the two phases. The pressure dependence of the free gas can be approximated by that of bulk methane while that of the adsorbed gas is more complex.

The quantity $N_{\mathrm{excess}}$ can be defined as the excess number of gas molecules in the pore space accommodated due to adsorption, compared to the number of bulk gas molecules that would have occupied the same pore volume in the absence of adsorption. This quantity is given as \cite{gensterblum2009european,gasparik2012high},
	
	\begin{equation}
		N_{\mathrm{excess}}=N_{G}-\rho_{\mathrm{free}}V_{\mathrm{pore}},
		\label{eq:Nexcess}
	\end{equation}
	
where $N_{G}$ is the total gas in the porous medium, $\rho_{\mathrm{free}}$ is the density of the bulk methane gas and $V_{\mathrm{pore}}$ is the volume of the pore space as specified by the manufacturer. The total gas in the pore space corresponds to the sum of the $\alpha$ and $\beta$ contributions of the $T_2$ distributions (Fig. \ref{fig:T2_distributions}). The second term $\rho_{\mathrm{free}}V_{\mathrm{pore}}$ corresponds to the gas in the pore space assuming no adsorption.

\begin{figure}
  	\includegraphics[width=1.1\linewidth]{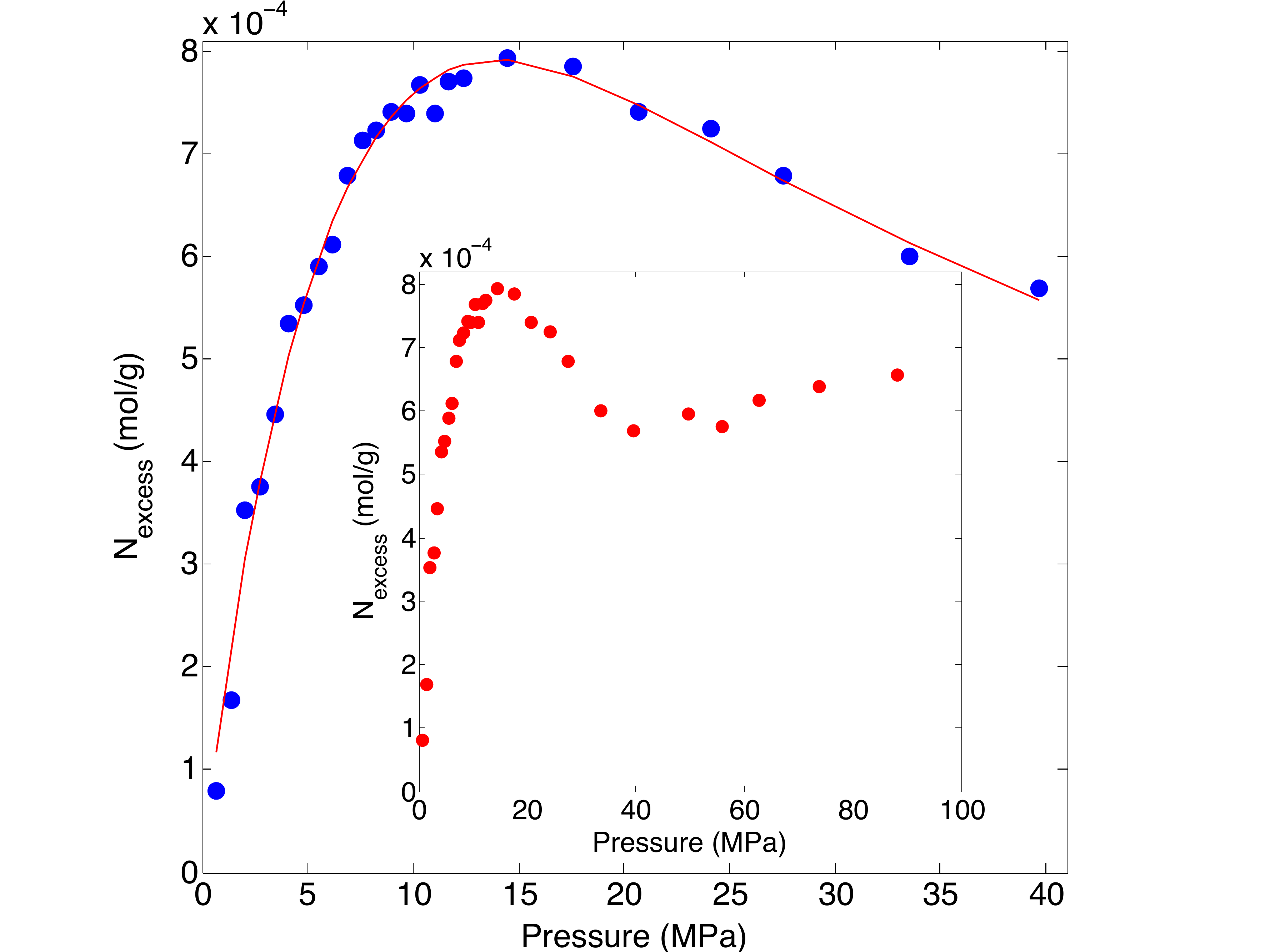}
 	 \caption{The quantity $N_{\mathrm{excess}}$ as a function of pressure determined experimentally is fit to the modified Langmuir isotherm (eq. \ref{eq:ModifiedLangmuir}) at the low pressure regime. This least squares fit with three degrees of freedom helps determine the adsorbed density layer, $\rho_{\mathrm{ads}}$, Langmuir pressure, $P_L$ and the maximum number of adsorbed molecules, $N_L$. The inset figure shows $N_{\mathrm{excess}}$ for the entire pressure regime, which can be described by multilayer adsorption models as described in the subsequent section. }
	 \label{fig:Langmuir_isotherm} 				 
\end{figure}

The excess gas in Vycor pore space may also be written as
\begin{equation}
N_{\mathrm{excess}}=N_{\mathrm{ads}}-\rho_{\mathrm{free}}V_{\mathrm{ads}}=N_{\mathrm{ads}}\Big(1-\frac{\rho_{\mathrm{free}}}{\rho_{\mathrm{ads}}}\Big),
\label{eq:Nexcess2}
\end{equation}
where $N_{\mathrm{ads}}$ is the number of adsorbed gas molecules, $\rho_{\mathrm{free}}$ and $\rho_{\mathrm{ads}}$ are the densities of the free and adsorbed gas phases and $V_{\mathrm{ads}}$ is the volume of the pore space occupied by the adsorbed gas. The pressure dependence of  $N_{\mathrm{ads}}$ helps identify the adsorption process due to its unique characteristics for monolayer \cite{langmuir1918adsorption} or multilayer adsorption \cite{brunauer1938adsorption} corresponding to the low or high pressure regimes. In the following sections we identify the pressure regimes corresponding to the Langmuir monolayer adsorption and BET multilayer adsorption and determine the density and volume occupied by the adsorbed layer, $\rho_{\mathrm{ads}}$ and $V_{\mathrm{ads}}$, as well as the number of adsorbed layers formed at higher pressures.

\subsubsection{Monolayer adsorption}
\label{Monolayer}
In the low pressure regime (0.7 Mpa$<P<$ 39.7 MPa) the adsorption process is established by a monolayer of adsorbed molecules which can be described by the Lamguir isotherm given as \cite{langmuir1918adsorption},

\begin{equation}
N_{\mathrm{ads}}=N_L\frac{P}{P+P_L},
\label{Langmuir}
\end{equation}
where $P$ is the applied pressure, $N_L$ the maximum number of adsorbed molecules at infinite pressure and $P_L$ the pressure at which half of the adsorption sites are occupied.

The resulting excess number of gas molecules can then be described by the modified Langmuir isotherm, obtained by combining eq. \ref{eq:Nexcess2} and eq. \ref{Langmuir} as \cite{gensterblum2009european},

\begin{equation}
	N_{\mathrm{excess}}=N_L\frac{P}{P+P_L}\Big(1-\frac{\rho_{\mathrm{free}}(P,T)}{\rho_{\mathrm{ads}}}\Big).
	\label{eq:ModifiedLangmuir}
\end{equation}
The $N_{\mathrm{excess}}$ computed using eq. \ref{eq:Nexcess} is shown in Fig. \ref{fig:Langmuir_isotherm} up to $P=39.6$ MPa where the Langmuir monolayer model still holds. The three unknown parameters $N_L$, $P_L$ and $\rho_{\mathrm{ads}}$ were obtained by performing least squares fit of the data to eq. \ref{eq:ModifiedLangmuir} with $\chi^2/dof=2.4$ \cite{bevington1969data,press2007numerical}. The values obtained by the fit for the Langmuir pressure, Langmuir volume and the equilibrium adsorbed gas density were, $P_L=10.80\pm 0.43$ [MPa], $N_L=(2.16\pm 0.06)\cdot10^{-4}[\mathrm{mol}\cdot \mathrm{g}^{-1}]$ and $\rho_{\mathrm{ads}}=(2.43\pm0.02)\cdot10^{-2}[\mathrm{mol}\cdot \mathrm{ml}^{-1}]$ and agree well with the values reported earlier for adsorption studies in activated carbon \cite{gensterblum2009european}. The density of the adsorbed layer ($\rho_{\mathrm{ads}}$) was determined to be 8.5\% lower than the density of liquid methane.

It can be seen in Fig. \ref{fig:Langmuir_isotherm} that the quantity $N_{\mathrm{excess}}$ increases at low pressures to reach a maximum at approximately $13.8$ MPa and then decreases at higher pressures. This is because of the increase in the number of adsorbed molecules at lower pressures ($P<13.8$ MPa), due to the presence of a large number of vacant surface sites. At pressures greater than $13.8$ MPa the number of adsorbed molecules slowly plateaus out, resulting in a slow decrease in $N_{\mathrm{excess}}$. At higher pressures and for monolayer adsorption, as the density of the free gas in the center of the pore approaches that of the adsorbed gas, $N_{\mathrm{excess}}$ should asymptotically approach zero. However, at pressure values of $P>39.6$ MPa we observe that the quantity $N_{\mathrm{excess}}$ starts to slowly increase again as shown in the inset of Fig. \ref{fig:Langmuir_isotherm}. This constitutes a deviation from the Langmuir isotherm behavior resulting from the formation of multiple adsorbed layers on the pore surface. In the following section we analyze the entire data set up to 89.7 MPa using a generalized isotherm, accounting for the excess number of molecules in the pore space due to multilayer adsorption.
 
\subsubsection{Multilayer adsorption}
\label{Multilayer}

In this section we consider the Brunauer-Emmett-Teller (BET) model which describes multilayer adsorption to describe the adsorption of methane in Vycor porous glass up to pressures of 89.7 MPa. The number of adsorbed molecules for multilayer adsorption process is given by \cite{brunauer1938adsorption},

\begin{equation}
N_{\mathrm{ads}}=N_L\frac{cx}{(1-x)}\Big( \frac{1-(N+1)x^N+Nx^{N+1}}{1+(c-1)x-cx^{N+1}}   \Big).
\label{BET}
\end{equation}
In the above equation, $x=P/P_0$, where $P_0$ is the saturation pressure, $N$ is the number of adsorbed layers, $N_L$ is the number of molecules in a monolayer and $c\propto e^{(E_1-E_N)/RT}$ is a unitless constant related to the heat energies of the first and $N$ layers ($E_1, E_N$). Substituting eq. \ref{BET} into eq. \ref{eq:Nexcess2} we obtain the modified BET isotherm given as,

\begin{equation}
\begin{split}
N_{\mathrm{excess}}=\frac{N_LcP}{P_0-P}\Bigg[ \frac{1-(N+1)(\frac{P}{P_0})^N+N(\frac{P}{P_0})^{N+1}}{1+(c-1)(\frac{P}{P_0})-c(\frac{P}{P_0})^{N+1}}\Bigg]\\ 
\cdot\Bigg( 1-\frac{\rho_f}{\rho_{\mathrm{ads}}} \Bigg).
\end{split}
\label{BETexcess}
\end{equation}
It should be noted that eq. \ref{BETexcess} reduces to a form of the Langmuir equation in the limit of $P<<P_0$ and $N=1$, given as,

\begin{equation}
N_{\mathrm{excess}}=N_L \frac{cP}{P_0+cP}\cdot\Bigg( 1-\frac{\rho_f}{\rho_{\mathrm{ads}}} \Bigg),
\label{BETLangmuir}
\end{equation}
which is equivalent to the Langmuir isotherm (eq. \ref{eq:ModifiedLangmuir}) for $P_L\equiv P_0/c$.

\begin{figure}
 	 \centering
  	\includegraphics[width=1.1\linewidth]{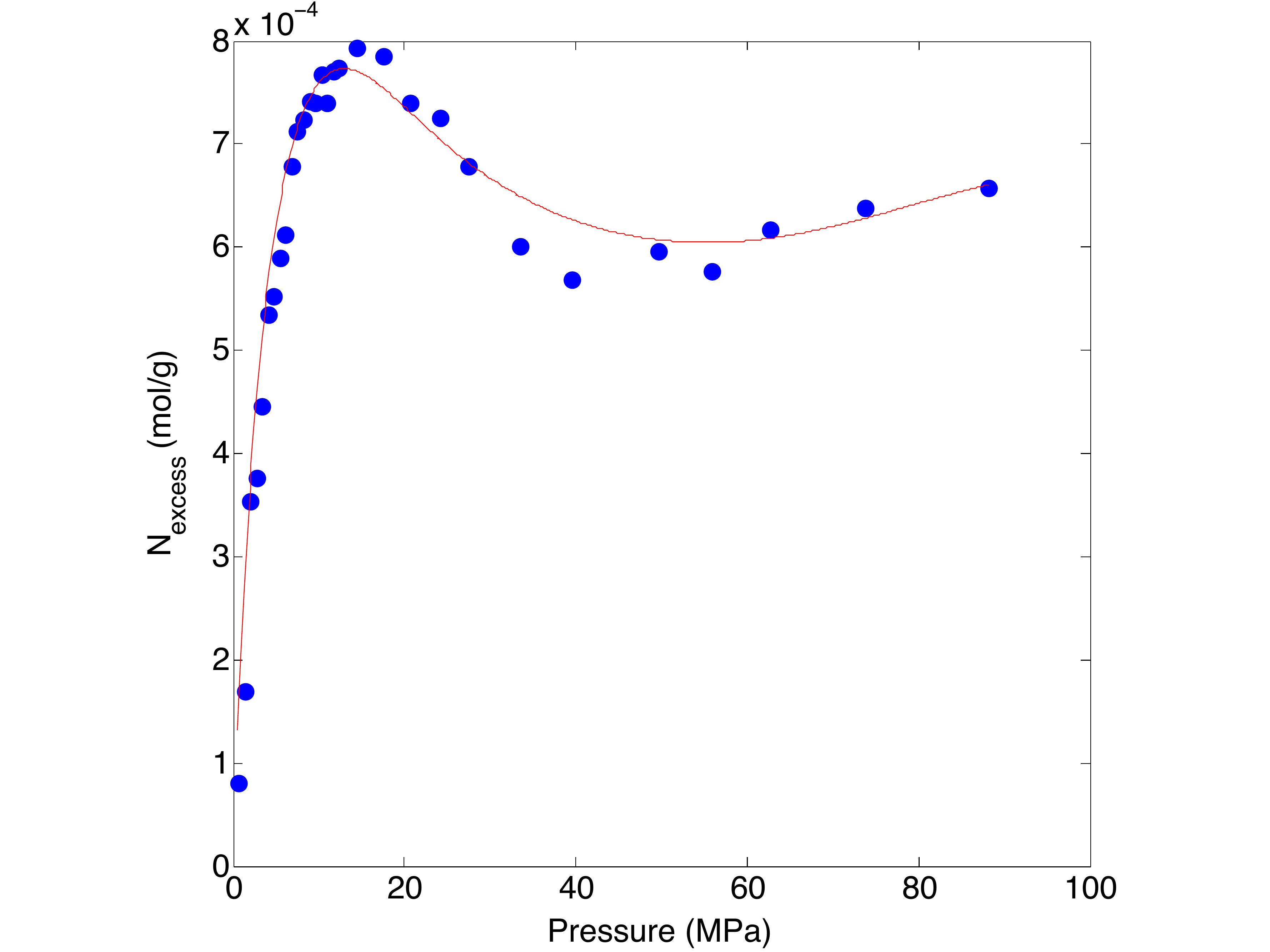}
 	 \caption{ The quantity $N_{\mathrm{excess}}$  is fit for the entire pressure range using the modified multilayer BET adsorption model (eq. \ref{BETexcess}).  This least squares fit with four degrees of freedom enables the determination of the adsorbed layers $N$, the constant $c$, the saturation pressure $P_0$ and the number of adsorbed molecules in a single layer $N_L$. The value for $\rho_{\mathrm{ads}}$ obtained from the monolayer Langmuir adsorption fit in the low pressure regime (fig 10) was used as an input in the fit.}
	 \label{fig:BET} 				 
\end{figure}

Performing nonlinear least squares fit of eq. \ref{BETexcess} for the excess gas in the pore space to the entire data set (shown in Fig. \ref{fig:BET}) up to pressures of 89.7 MPa we obtain, $P_0=168.0\pm14.1$ [MPa], $N_L=(1.4\pm0.3)\cdot10^{-4}[\mathrm{mol}\cdot \mathrm{g}^{-1}]$, $c=(21.6\pm4.6)$ , $N=(9.7\pm3.1)$. The fit had $\chi^2/dof=0.7$ \cite{press2007numerical,bevington1969data}. The value for $\rho_{\mathrm{ads}}$ obtained for the monolayer Langmuir adsorption in the low pressure regime (see sec. \ref{Monolayer}) was used as a constant in eq. \ref{BETLangmuir} during least squares minimization in order to reduce the complexity of the fit.

The constant $c$ obtained by the fit results in a $(E_1 - E_N) = 7.5kJ/mol$ energy gap between the adsorbed layers at $T=22^{\circ}C$, where $E_1$ is the activation energy of adsorption. This value agrees with the activation energy values reported earlier for methane adsorption on activated carbon \cite{himeno2005high}. The saturation pressure $P_0$ and constant $c$ obtained by fitting the experimental data help determine the Langmuir pressure ($P_L=P_0/c$). The fit value for the Langmuir pressure is in good agreement with that obtained by the monolayer Langmuir model for the low pressure regime (sec. \ref{Monolayer}-Monolayer adsorption). The number of layers of methane with average molecule diameter of 3.9 $\mathrm{\AA}$ that can be formed into the Vycor pore with an average diameter of 5.7 nm is approximately 7.4. Note that the fit overestimates the number of adsorbed layers ($N=(9.7\pm3.1)$) formed in the Vycor pore but the obtained theoretical value is within the experimental uncertainty. 

\subsubsection{Separation of free and adsorbed gas}
\label{Sep}
 
 To separate the gas in the Vycor nanopores into free and adsorbed fractions for the entire pressure range from 0.7 MPa-89.7 MPa we make use of the density of the adsorbed layer obtained from the Langmuir isotherm fit. At low pressures ($P<39.6$ MPa) the $N_{\mathrm{excess}}$ can be alternatively written in terms of the density of the adsorbed layer as,
 
 \begin{equation}
N_{\mathrm{excess}}\mid_{P=39.6\mathrm{MPa}}=\rho_{\mathrm{ads}}V_{\mathrm{ads}}-\rho_{\mathrm{bulk}}\mid_{P=39.6\mathrm{MPa}}V_{\mathrm{ads}}.
\label{eq: Vads}
\end{equation}

 \begin{figure}
 	 \centering
  	\includegraphics[width=1.1\linewidth]{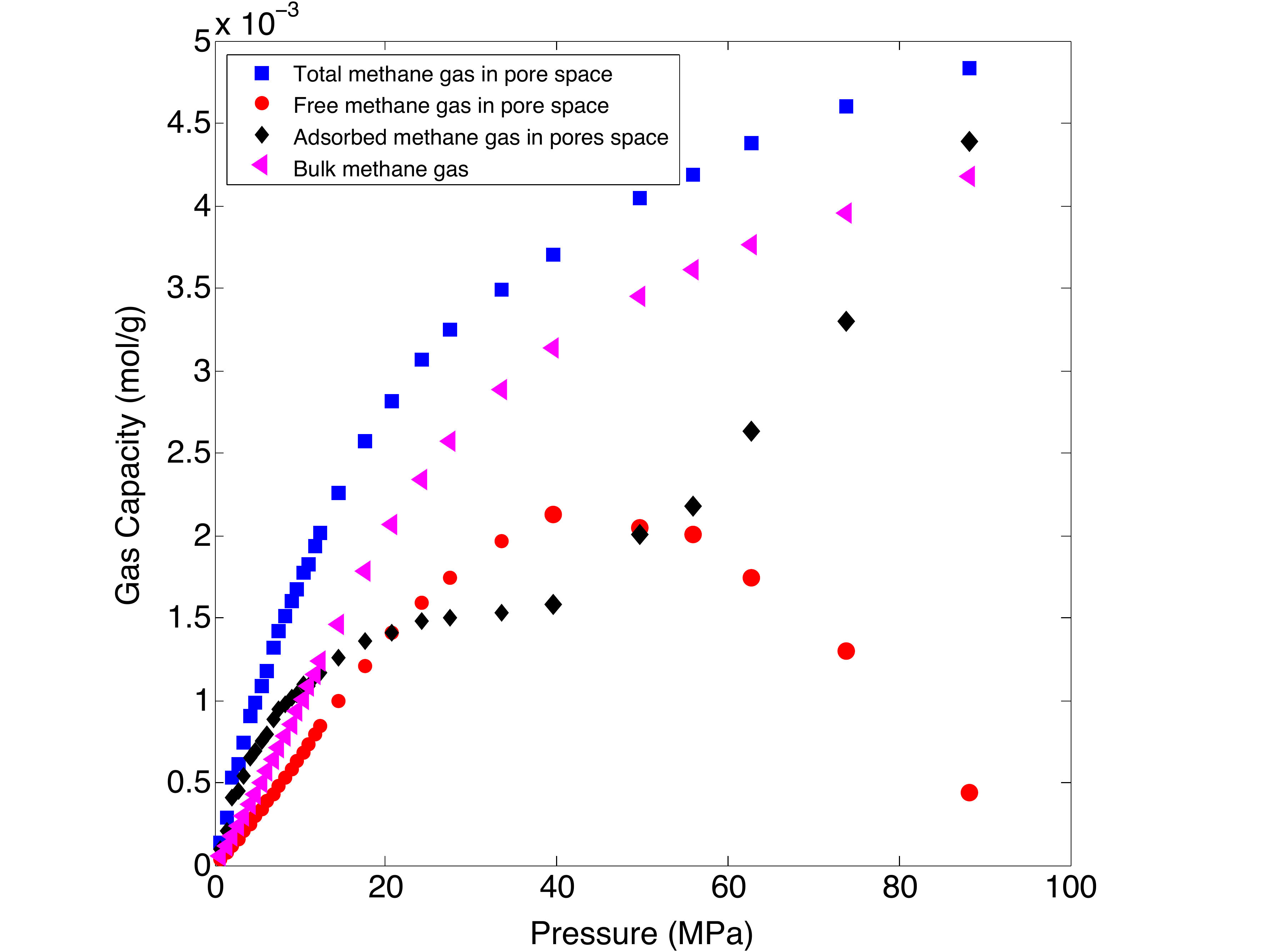}
 	 \caption{Separation of the total gas into free and adsorbed. The blue squares correspond to $^{1}H$ inside Vycor. Based on the $V_{\mathrm{ads}}$ calculated from eq. \ref{eq: Vads}, the total number of adsorbed $^{1}H$ was calculated (black diamonds). The red circles correspond to the total number of free $^{1}H$ in the pores. The bulk methane gas in Vycor (assuming no adsorption) is also shown in magenta triangles.  } 
	 \label{fig:key_product}				 
\end{figure} 

 From the above equation the volume of a monolayer of adsorbed gas ($V_{\mathrm{ads}}$) can be determined using the $\rho_{\mathrm{ads}}$ obtained from the Langmuir fit and the known values of $\rho_{\mathrm{bulk}}$. Using the adsorbed gas volume together with the pore volume of Vycor, the gas stored in the Vycor nanopores can be separated into free and adsorbed gas on the pore walls, for pressures up to $39.6$ MPa. For pressures higher than $39.6$ MPa, $V_{\mathrm{ads}}$ is pressure dependent as multiple layers start to form and is therefore determined as,
\begin{equation}
V_{\mathrm{ads}}(P)\mid^{P=89.7\mathrm{MPa}}_{\tiny{P=39.6\mathrm{MPa}}}=\frac{N_{\mathrm{excess}}(P)\mid^{P=89.7\mathrm{MPa}}_{\tiny{P=39.6\mathrm{MPa}}}}{\rho_{\mathrm{ads}}-\rho_{\mathrm{bulk}}}.
\end{equation} 
 
 The results of the free, adsorbed and total methane gas in Vycor porous glass are shown in Fig \ref{fig:key_product}. Up to a pressure of $39.6$ MPa the ``free gas'' molecules in the pore interiors (red circles in Fig \ref{fig:key_product}) show a bulk-like pressure dependence while the adsorbed gas fraction increases faster at lower pressures but slowly saturates at higher pressures. This indicates that at low pressures ($P<12$ MPa) more gas is stored in the pore space as adsorbed gas. As discussed in Sec. \ref{Multilayer}, multiple adsorbed layers start to form for pressures $P>39.6$ MPa and therefore a rapid drop is observed in the ``free gas" fraction. On the other hand the adsorbed gas in the pore space shows a rapid increase up to $P=89.7$ MPa, where it is approximately 91\% of the total methane gas stored in the pore space. The amount of bulk methane gas that could be accommodated in the pore volume, in the absence of adsorption is also shown in magenta. This way the total gas stored in the Vycor nanopores can be separated into free and adsorbed fractions in the entire pressure range from 0.7 Mpa to 89.7 MPa using high field NMR relaxometry.
 
\section{Conclusions}
   
 In the present work we demonstrate the application of high field NMR for characterization of the storage properties of methane gas in nanometer sized pore spaces. We carry out laboratory experiments at a wide pressure range (0.7 MPa to 89.7 MPa), that covers most conditions encountered in naturally existing shale gas and tight gas reservoirs. We demonstrate that the relaxation of the bulk methane gas, which forms the dominant fraction of natural gas, is dominated by the spin rotation Hamiltonian up to the highest studied pressures of 89.7 MPa. The sensitivity of the $T_2$ relaxation distributions to the slow motions enable the separation of the gas molecules in the pore space and thereby aid gas storage and HI measurements as a function of pressure.
 
 The quantity of methane gas in Vycor pore space was used to determine the excess gas due to adsorption on the pore walls. Two pressure regimes that correspond to monolayer and multilayer adsorption were identified. In the low pressure regime from $0.7<P<39.6$ MPa the excess gas was fit to the Langmuir isotherm and the Langmuir pressure, volume and equilibrium density of the adsorbed layer were determined. The adsorbed gas density was discovered to be 8.5\% lower than that of liquid methane. In the high pressure regime from 39.6 Mpa to 89.7 MPa the pressure dependence of the excess gas shows evidence of multilayer adsorption and was modeled using BET theory to determine the number of adsorbed layers formed. Using the adsorbed gas density, the gas in the Vycor pore space was separated into free and adsorbed fractions for the entire pressure range. The adsorbed gas fraction was shown to increase sharply for pressures above $39.6$ MPa due the formation of multiple adsorbed layers, while the quantity of free gas molecules in the pore space decreased correspondingly. 
   
Vycor porous glass with 5.7 nm pore size used in these studies can be considered as a model system to understand storage properties of natural gas in shale gas and tight gas rock formations. While the pore size of the Vycor porous glass is comparable to the small pores in the shale rock samples, important differences include the low porosities, permeability and also the presence of paramagnetic impurities in the naturally occurring shale rocks. The presence of internal gradients due to paramagnetic impurities in the shale nanopores are averaged out by the fast diffusing gas molecules, but can affect the annulus gas outside the rock resulting in their overlap in the relaxation distributions. The low porosity and permeability of shale rocks also make the NMR experiments time consuming in order to achieve acceptable signal to noise ratio and complete gas saturation. In spite of these challenges, the sensitivity and precision of the high field NMR relaxometry experiments make it a viable methodology for studying the gas storage and transport characteristics of natural gas in such micro and mesoporous media.\\

\section{Acknowledgements}
AP was funded through the Schlumberger-Doll Research internship program during the course of this work. The authors thank Schlumberger for permission to publish the results and Yi-Qiao Song, Martin Hurlimann, Gregory S. Boutis and Steven W. Morgan for useful discussions regarding the experimental findings.

\bibliography{Langmuir}

\begin{thebibliography}{45}%
\makeatletter
\providecommand \@ifxundefined [1]{%
 \@ifx{#1\undefined}
}%
\providecommand \@ifnum [1]{%
 \ifnum #1\expandafter \@firstoftwo
 \else \expandafter \@secondoftwo
 \fi
}%
\providecommand \@ifx [1]{%
 \ifx #1\expandafter \@firstoftwo
 \else \expandafter \@secondoftwo
 \fi
}%
\providecommand \natexlab [1]{#1}%
\providecommand \enquote  [1]{``#1''}%
\providecommand \bibnamefont  [1]{#1}%
\providecommand \bibfnamefont [1]{#1}%
\providecommand \citenamefont [1]{#1}%
\providecommand \href@noop [0]{\@secondoftwo}%
\providecommand \href [0]{\begingroup \@sanitize@url \@href}%
\providecommand \@href[1]{\@@startlink{#1}\@@href}%
\providecommand \@@href[1]{\endgroup#1\@@endlink}%
\providecommand \@sanitize@url [0]{\catcode `\\12\catcode `\$12\catcode
  `\&12\catcode `\#12\catcode `\^12\catcode `\_12\catcode `\%12\relax}%
\providecommand \@@startlink[1]{}%
\providecommand \@@endlink[0]{}%
\providecommand \url  [0]{\begingroup\@sanitize@url \@url }%
\providecommand \@url [1]{\endgroup\@href {#1}{\urlprefix }}%
\providecommand \urlprefix  [0]{URL }%
\providecommand \Eprint [0]{\href }%
\providecommand \doibase [0]{http://dx.doi.org/}%
\providecommand \selectlanguage [0]{\@gobble}%
\providecommand \bibinfo  [0]{\@secondoftwo}%
\providecommand \bibfield  [0]{\@secondoftwo}%
\providecommand \translation [1]{[#1]}%
\providecommand \BibitemOpen [0]{}%
\providecommand \bibitemStop [0]{}%
\providecommand \bibitemNoStop [0]{.\EOS\space}%
\providecommand \EOS [0]{\spacefactor3000\relax}%
\providecommand \BibitemShut  [1]{\csname bibitem#1\endcsname}%
\let\auto@bib@innerbib\@empty
\bibitem [{\citenamefont {Mair}\ \emph {et~al.}(1999)\citenamefont {Mair},
  \citenamefont {Wong}, \citenamefont {Hoffmann}, \citenamefont {H\"urlimann},
  \citenamefont {Patz}, \citenamefont {Schwartz},\ and\ \citenamefont
  {Walsworth}}]{mair1999probing}%
  \BibitemOpen
  \bibfield  {author} {\bibinfo {author} {\bibfnamefont {R.~W.}\ \bibnamefont
  {Mair}}, \bibinfo {author} {\bibfnamefont {G.~P.}\ \bibnamefont {Wong}},
  \bibinfo {author} {\bibfnamefont {D.}~\bibnamefont {Hoffmann}}, \bibinfo
  {author} {\bibfnamefont {M.~D.}\ \bibnamefont {H\"urlimann}}, \bibinfo
  {author} {\bibfnamefont {S.}~\bibnamefont {Patz}}, \bibinfo {author}
  {\bibfnamefont {L.~M.}\ \bibnamefont {Schwartz}}, \ and\ \bibinfo {author}
  {\bibfnamefont {R.~L.}\ \bibnamefont {Walsworth}},\ }\bibfield  {title}
  {\enquote {\bibinfo {title} {Probing porous media with gas diffusion
  {NMR}},}\ }\href {\doibase 10.1103/PhysRevLett.83.3324} {\bibfield  {journal}
  {\bibinfo  {journal} {Phys. Rev. Lett.}\ }\textbf {\bibinfo {volume} {83}},\
  \bibinfo {pages} {3324--3327} (\bibinfo {year} {1999})}\BibitemShut {NoStop}%
\bibitem [{\citenamefont {Boutis}\ \emph {et~al.}(2007)\citenamefont {Boutis},
  \citenamefont {Renner}, \citenamefont {Isahkarov}, \citenamefont {Islam},
  \citenamefont {Kannangara}, \citenamefont {Kaur}, \citenamefont {Mananga},
  \citenamefont {Ntekim}, \citenamefont {Rumala},\ and\ \citenamefont
  {Wei}}]{boutis2007high}%
  \BibitemOpen
  \bibfield  {author} {\bibinfo {author} {\bibfnamefont {G.~S.}\ \bibnamefont
  {Boutis}}, \bibinfo {author} {\bibfnamefont {C.}~\bibnamefont {Renner}},
  \bibinfo {author} {\bibfnamefont {T.}~\bibnamefont {Isahkarov}}, \bibinfo
  {author} {\bibfnamefont {T.}~\bibnamefont {Islam}}, \bibinfo {author}
  {\bibfnamefont {L.}~\bibnamefont {Kannangara}}, \bibinfo {author}
  {\bibfnamefont {P.}~\bibnamefont {Kaur}}, \bibinfo {author} {\bibfnamefont
  {E.}~\bibnamefont {Mananga}}, \bibinfo {author} {\bibfnamefont
  {A.}~\bibnamefont {Ntekim}}, \bibinfo {author} {\bibfnamefont
  {Y.}~\bibnamefont {Rumala}}, \ and\ \bibinfo {author} {\bibfnamefont
  {D.}~\bibnamefont {Wei}},\ }\bibfield  {title} {\enquote {\bibinfo {title}
  {High resolution q-space imaging studies of water in elastin},}\ }\href
  {http://onlinelibrary.wiley.com/doi/10.1002/bip.20838/full} {\bibfield
  {journal} {\bibinfo  {journal} {Biopolymers}\ }\textbf {\bibinfo {volume}
  {87}},\ \bibinfo {pages} {352--359} (\bibinfo {year} {2007})}\BibitemShut
  {NoStop}%
\bibitem [{\citenamefont {Novikov}\ \emph {et~al.}(2014)\citenamefont
  {Novikov}, \citenamefont {Jensen}, \citenamefont {Helpern},\ and\
  \citenamefont {Fieremans}}]{novikov2012characterizing}%
  \BibitemOpen
  \bibfield  {author} {\bibinfo {author} {\bibfnamefont {D.~S.}\ \bibnamefont
  {Novikov}}, \bibinfo {author} {\bibfnamefont {J.~H.}\ \bibnamefont {Jensen}},
  \bibinfo {author} {\bibfnamefont {J.~A.}\ \bibnamefont {Helpern}}, \ and\
  \bibinfo {author} {\bibfnamefont {E.}~\bibnamefont {Fieremans}},\ }\bibfield
  {title} {\enquote {\bibinfo {title} {Revealing mesoscopic structural
  universality with diffusion},}\ }\href
  {http://www.pnas.org/content/111/14/5088.short} {\bibfield  {journal}
  {\bibinfo  {journal} {Proc. Natl. Acad. Sci. U.S.A.}\ }\textbf {\bibinfo
  {volume} {111}},\ \bibinfo {pages} {5088--5093} (\bibinfo {year}
  {2014})}\BibitemShut {NoStop}%
\bibitem [{\citenamefont {Mitra}\ \emph {et~al.}(1992)\citenamefont {Mitra},
  \citenamefont {Sen}, \citenamefont {Schwartz},\ and\ \citenamefont
  {Le~Doussal}}]{mitra1992diffusion}%
  \BibitemOpen
  \bibfield  {author} {\bibinfo {author} {\bibfnamefont {P.~P.}\ \bibnamefont
  {Mitra}}, \bibinfo {author} {\bibfnamefont {P.~N.}\ \bibnamefont {Sen}},
  \bibinfo {author} {\bibfnamefont {L.~M.}\ \bibnamefont {Schwartz}}, \ and\
  \bibinfo {author} {\bibfnamefont {P.}~\bibnamefont {Le~Doussal}},\ }\bibfield
   {title} {\enquote {\bibinfo {title} {Diffusion propagator as a probe of the
  structure of porous media},}\ }\href
  {http://journals.aps.org/prl/abstract/10.1103/PhysRevLett.68.3555} {\bibfield
   {journal} {\bibinfo  {journal} {Phys. Rev. Lett.}\ }\textbf {\bibinfo
  {volume} {68}},\ \bibinfo {pages} {3555--3558} (\bibinfo {year}
  {1992})}\BibitemShut {NoStop}%
\bibitem [{\citenamefont {Callaghan}\ \emph {et~al.}(1991)\citenamefont
  {Callaghan}, \citenamefont {Coy}, \citenamefont {MacGowan}, \citenamefont
  {Packer},\ and\ \citenamefont {Zelaya}}]{diffraction}%
  \BibitemOpen
  \bibfield  {author} {\bibinfo {author} {\bibfnamefont {P.~T.}\ \bibnamefont
  {Callaghan}}, \bibinfo {author} {\bibfnamefont {A.}~\bibnamefont {Coy}},
  \bibinfo {author} {\bibfnamefont {D.}~\bibnamefont {MacGowan}}, \bibinfo
  {author} {\bibfnamefont {K.~J.}\ \bibnamefont {Packer}}, \ and\ \bibinfo
  {author} {\bibfnamefont {F.~O.}\ \bibnamefont {Zelaya}},\ }\bibfield  {title}
  {\enquote {\bibinfo {title} {Diffraction-like effects in {NMR} diffusion
  studies of fluids in porous solids},}\ }\href
  {http://dx.doi.org/10.1038/351467a0} {\bibfield  {journal} {\bibinfo
  {journal} {Nature}\ }\textbf {\bibinfo {volume} {351}},\ \bibinfo {pages}
  {467--469} (\bibinfo {year} {1991})}\BibitemShut {NoStop}%
\bibitem [{\citenamefont {Wang}\ \emph {et~al.}(2014)\citenamefont {Wang},
  \citenamefont {Mutina},\ and\ \citenamefont {Kausik}}]{wang2014high}%
  \BibitemOpen
  \bibfield  {author} {\bibinfo {author} {\bibfnamefont {H.-J.}\ \bibnamefont
  {Wang}}, \bibinfo {author} {\bibfnamefont {A.}~\bibnamefont {Mutina}}, \ and\
  \bibinfo {author} {\bibfnamefont {R.}~\bibnamefont {Kausik}},\ }\bibfield
  {title} {\enquote {\bibinfo {title} {High-field nuclear magnetic resonance
  observation of gas shale fracturing by methane gas},}\ }\href
  {http://pubs.acs.org/doi/abs/10.1021/ef5002937} {\bibfield  {journal}
  {\bibinfo  {journal} {Energy \& Fuels}\ }\textbf {\bibinfo {volume} {28}},\
  \bibinfo {pages} {3638--3644} (\bibinfo {year} {2014})}\BibitemShut {NoStop}%
\bibitem [{\citenamefont {Gasparik}\ \emph {et~al.}(2012)\citenamefont
  {Gasparik}, \citenamefont {Ghanizadeh}, \citenamefont {Bertier},
  \citenamefont {Gensterblum}, \citenamefont {Bouw},\ and\ \citenamefont
  {Krooss}}]{gasparik2012high}%
  \BibitemOpen
  \bibfield  {author} {\bibinfo {author} {\bibfnamefont {M.}~\bibnamefont
  {Gasparik}}, \bibinfo {author} {\bibfnamefont {A.}~\bibnamefont
  {Ghanizadeh}}, \bibinfo {author} {\bibfnamefont {P.}~\bibnamefont {Bertier}},
  \bibinfo {author} {\bibfnamefont {Y.}~\bibnamefont {Gensterblum}}, \bibinfo
  {author} {\bibfnamefont {S.}~\bibnamefont {Bouw}}, \ and\ \bibinfo {author}
  {\bibfnamefont {B.~M.}\ \bibnamefont {Krooss}},\ }\bibfield  {title}
  {\enquote {\bibinfo {title} {High-pressure methane sorption isotherms of
  black shales from the netherlands},}\ }\href
  {http://pubs.acs.org/doi/abs/10.1021/ef300405g} {\bibfield  {journal}
  {\bibinfo  {journal} {Energy \& Fuels}\ }\textbf {\bibinfo {volume} {26}},\
  \bibinfo {pages} {4995--5004} (\bibinfo {year} {2012})}\BibitemShut {NoStop}%
\bibitem [{\citenamefont {Kausik}\ \emph {et~al.}(2011)\citenamefont {Kausik},
  \citenamefont {Minh}, \citenamefont {Zielinski}, \citenamefont
  {Vissapragada}, \citenamefont {Akkurt}, \citenamefont {Song}, \citenamefont
  {Liu}, \citenamefont {Jones},\ and\ \citenamefont
  {Blair}}]{kausik2011characterization}%
  \BibitemOpen
  \bibfield  {author} {\bibinfo {author} {\bibfnamefont {R.}~\bibnamefont
  {Kausik}}, \bibinfo {author} {\bibfnamefont {C.~C.}\ \bibnamefont {Minh}},
  \bibinfo {author} {\bibfnamefont {L.}~\bibnamefont {Zielinski}}, \bibinfo
  {author} {\bibfnamefont {B.}~\bibnamefont {Vissapragada}}, \bibinfo {author}
  {\bibfnamefont {R.}~\bibnamefont {Akkurt}}, \bibinfo {author} {\bibfnamefont
  {Y.}~\bibnamefont {Song}}, \bibinfo {author} {\bibfnamefont {C.}~\bibnamefont
  {Liu}}, \bibinfo {author} {\bibfnamefont {S.}~\bibnamefont {Jones}}, \ and\
  \bibinfo {author} {\bibfnamefont {E.}~\bibnamefont {Blair}},\ }\bibfield
  {title} {\enquote {\bibinfo {title} {Characterization of gas dynamics in
  kerogen nanopores by {NMR}},}\ }in\ \href
  {https://www.onepetro.org/conference-paper/SPE-147198-MS} {\emph {\bibinfo
  {booktitle} {Proceedings of the SPE Annual Technical Conference and
  Exhibition}}}\ (\bibinfo {year} {2011})\BibitemShut {NoStop}%
\bibitem [{\citenamefont {Panella}\ \emph {et~al.}(2005)\citenamefont
  {Panella}, \citenamefont {Hirscher},\ and\ \citenamefont
  {Roth}}]{panella2005hydrogen}%
  \BibitemOpen
  \bibfield  {author} {\bibinfo {author} {\bibfnamefont {B.}~\bibnamefont
  {Panella}}, \bibinfo {author} {\bibfnamefont {M.}~\bibnamefont {Hirscher}}, \
  and\ \bibinfo {author} {\bibfnamefont {S.}~\bibnamefont {Roth}},\ }\bibfield
  {title} {\enquote {\bibinfo {title} {Hydrogen adsorption in different carbon
  nanostructures},}\ }\href
  {http://www.sciencedirect.com/science/article/pii/S000862230500206X}
  {\bibfield  {journal} {\bibinfo  {journal} {Carbon}\ }\textbf {\bibinfo
  {volume} {43}},\ \bibinfo {pages} {2209--2214} (\bibinfo {year}
  {2005})}\BibitemShut {NoStop}%
\bibitem [{\citenamefont {Horch}\ \emph {et~al.}(2014)\citenamefont {Horch},
  \citenamefont {Schlayer},\ and\ \citenamefont {Stallmach}}]{horch2014high}%
  \BibitemOpen
  \bibfield  {author} {\bibinfo {author} {\bibfnamefont {C.}~\bibnamefont
  {Horch}}, \bibinfo {author} {\bibfnamefont {S.}~\bibnamefont {Schlayer}}, \
  and\ \bibinfo {author} {\bibfnamefont {F.}~\bibnamefont {Stallmach}},\
  }\bibfield  {title} {\enquote {\bibinfo {title} {High-pressure low-field 1{H}
  {NMR} relaxometry in nanoporous materials},}\ }\href {\doibase
  http://dx.doi.org/10.1016/j.jmr.2014.01.002} {\bibfield  {journal} {\bibinfo
  {journal} {J. Magn. Reson.}\ }\textbf {\bibinfo {volume} {240}},\ \bibinfo
  {pages} {24 -- 33} (\bibinfo {year} {2014})}\BibitemShut {NoStop}%
\bibitem [{\citenamefont {Bear}(2013)}]{bear2013dynamics}%
  \BibitemOpen
  \bibfield  {author} {\bibinfo {author} {\bibfnamefont {J.}~\bibnamefont
  {Bear}},\ }\href@noop {} {\emph {\bibinfo {title} {Dynamics of Fluids in
  Porous Media}}}\ (\bibinfo  {publisher} {Dover Publications},\ \bibinfo
  {year} {2013})\BibitemShut {NoStop}%
\bibitem [{\citenamefont {Dvoyashkin}\ \emph {et~al.}(2007)\citenamefont
  {Dvoyashkin}, \citenamefont {Valiullin},\ and\ \citenamefont
  {K{\"a}rger}}]{dvoyashkin2007supercritical}%
  \BibitemOpen
  \bibfield  {author} {\bibinfo {author} {\bibfnamefont {M.}~\bibnamefont
  {Dvoyashkin}}, \bibinfo {author} {\bibfnamefont {R.}~\bibnamefont
  {Valiullin}}, \ and\ \bibinfo {author} {\bibfnamefont {J.}~\bibnamefont
  {K{\"a}rger}},\ }\bibfield  {title} {\enquote {\bibinfo {title}
  {Supercritical fluids in mesopores-new insight using {NMR}},}\ }\href
  {http://link.springer.com/article/10.1007/s10450-007-9064-y} {\bibfield
  {journal} {\bibinfo  {journal} {Adsorption}\ }\textbf {\bibinfo {volume}
  {13}},\ \bibinfo {pages} {197--200} (\bibinfo {year} {2007})}\BibitemShut
  {NoStop}%
\bibitem [{\citenamefont {Valiullin}\ and\ \citenamefont
  {K{\"a}rger}(2011)}]{valiullin2011impact}%
  \BibitemOpen
  \bibfield  {author} {\bibinfo {author} {\bibfnamefont {R.}~\bibnamefont
  {Valiullin}}\ and\ \bibinfo {author} {\bibfnamefont {J.}~\bibnamefont
  {K{\"a}rger}},\ }\bibfield  {title} {\enquote {\bibinfo {title} {The impact
  of mesopores on mass transfer in nanoporous materials: evidence of diffusion
  measurement by {NMR}},}\ }\href
  {http://onlinelibrary.wiley.com/doi/10.1002/cite.201000208/abstract}
  {\bibfield  {journal} {\bibinfo  {journal} {Chem. Ing. Tech.}\ }\textbf
  {\bibinfo {volume} {83}},\ \bibinfo {pages} {166--176} (\bibinfo {year}
  {2011})}\BibitemShut {NoStop}%
\bibitem [{\citenamefont {Valiullin}\ \emph {et~al.}(2009)\citenamefont
  {Valiullin}, \citenamefont {K{\"a}rger},\ and\ \citenamefont
  {Gl{\"a}ser}}]{valiullin2009correlating}%
  \BibitemOpen
  \bibfield  {author} {\bibinfo {author} {\bibfnamefont {R.}~\bibnamefont
  {Valiullin}}, \bibinfo {author} {\bibfnamefont {J.}~\bibnamefont
  {K{\"a}rger}}, \ and\ \bibinfo {author} {\bibfnamefont {R.}~\bibnamefont
  {Gl{\"a}ser}},\ }\bibfield  {title} {\enquote {\bibinfo {title} {Correlating
  phase behaviour and diffusion in mesopores: perspectives revealed by pulsed
  field gradient {NMR}},}\ }\href
  {http://pubs.rsc.org/en/Content/ArticleLanding/2009/CP/b822939b#!divAbstract}
  {\bibfield  {journal} {\bibinfo  {journal} {Phys. Chem. Chem. Phys.}\
  }\textbf {\bibinfo {volume} {11}},\ \bibinfo {pages} {2833--2853} (\bibinfo
  {year} {2009})}\BibitemShut {NoStop}%
\bibitem [{\citenamefont {Riehl}\ and\ \citenamefont
  {Koch}(1972)}]{riehl1972nmr}%
  \BibitemOpen
  \bibfield  {author} {\bibinfo {author} {\bibfnamefont {J.}~\bibnamefont
  {Riehl}}\ and\ \bibinfo {author} {\bibfnamefont {K.}~\bibnamefont {Koch}},\
  }\bibfield  {title} {\enquote {\bibinfo {title} {{NMR} relaxation of adsorbed
  gases: Methane on graphite},}\ }\href
  {http://scitation.aip.org/content/aip/journal/jcp/57/5/10.1063/1.1678553}
  {\bibfield  {journal} {\bibinfo  {journal} {J. Chem. Phys.}\ }\textbf
  {\bibinfo {volume} {57}},\ \bibinfo {pages} {2199--2208} (\bibinfo {year}
  {1972})}\BibitemShut {NoStop}%
\bibitem [{\citenamefont {Carr}\ and\ \citenamefont
  {Purcell}(1954)}]{carr1954effects}%
  \BibitemOpen
  \bibfield  {author} {\bibinfo {author} {\bibfnamefont {H.~Y.}\ \bibnamefont
  {Carr}}\ and\ \bibinfo {author} {\bibfnamefont {E.~M.}\ \bibnamefont
  {Purcell}},\ }\bibfield  {title} {\enquote {\bibinfo {title} {Effects of
  diffusion on free precession in nuclear magnetic resonance experiments},}\
  }\href {http://journals.aps.org/pr/abstract/10.1103/PhysRev.94.630}
  {\bibfield  {journal} {\bibinfo  {journal} {Phys. Rev.}\ }\textbf {\bibinfo
  {volume} {94}},\ \bibinfo {pages} {630} (\bibinfo {year} {1954})}\BibitemShut
  {NoStop}%
\bibitem [{\citenamefont {Meiboom}\ and\ \citenamefont
  {Gill}(1958)}]{meiboom1958modified}%
  \BibitemOpen
  \bibfield  {author} {\bibinfo {author} {\bibfnamefont {S.}~\bibnamefont
  {Meiboom}}\ and\ \bibinfo {author} {\bibfnamefont {D.}~\bibnamefont {Gill}},\
  }\bibfield  {title} {\enquote {\bibinfo {title} {Modified spin-echo method
  for measuring nuclear relaxation times},}\ }\href
  {http://scitation.aip.org/content/aip/journal/rsi/29/8/10.1063/1.1716296}
  {\bibfield  {journal} {\bibinfo  {journal} {Rev. Sci. Instrum.}\ }\textbf
  {\bibinfo {volume} {29}},\ \bibinfo {pages} {688--691} (\bibinfo {year}
  {1958})}\BibitemShut {NoStop}%
\bibitem [{\citenamefont {Hurlimann}\ \emph {et~al.}(1994)\citenamefont
  {Hurlimann}, \citenamefont {Helmer}, \citenamefont {Latour},\ and\
  \citenamefont {Sotak}}]{hurlimann1994restricted}%
  \BibitemOpen
  \bibfield  {author} {\bibinfo {author} {\bibfnamefont {M.}~\bibnamefont
  {Hurlimann}}, \bibinfo {author} {\bibfnamefont {K.~G.}\ \bibnamefont
  {Helmer}}, \bibinfo {author} {\bibfnamefont {L.}~\bibnamefont {Latour}}, \
  and\ \bibinfo {author} {\bibfnamefont {C.~H.}\ \bibnamefont {Sotak}},\
  }\bibfield  {title} {\enquote {\bibinfo {title} {Restricted diffusion in
  sedimentary rocks. determination of surface-area-to-volume ratio and surface
  relaxivity},}\ }\href
  {http://www.sciencedirect.com/science/article/pii/S1064185884712435}
  {\bibfield  {journal} {\bibinfo  {journal} {J. Magn. Reson., Ser. A}\
  }\textbf {\bibinfo {volume} {111}},\ \bibinfo {pages} {169--178} (\bibinfo
  {year} {1994})}\BibitemShut {NoStop}%
\bibitem [{\citenamefont {Linstrom}\ and\ \citenamefont
  {Mallard}(2001)}]{linstrom2001nist}%
  \BibitemOpen
  \bibfield  {author} {\bibinfo {author} {\bibfnamefont {P.~J.}\ \bibnamefont
  {Linstrom}}\ and\ \bibinfo {author} {\bibfnamefont {W.~G.}\ \bibnamefont
  {Mallard}},\ }\bibfield  {title} {\enquote {\bibinfo {title} {The {NIST}
  chemistry webbook: A chemical data resource on the internet},}\ }\href
  {http://pubs.acs.org/doi/abs/10.1021/je000236i} {\bibfield  {journal}
  {\bibinfo  {journal} {J. Chem. Eng. Data}\ }\textbf {\bibinfo {volume}
  {46}},\ \bibinfo {pages} {1059--1063} (\bibinfo {year} {2001})}\BibitemShut
  {NoStop}%
\bibitem [{\citenamefont {Hari}\ \emph {et~al.}(1998)\citenamefont {Hari},
  \citenamefont {Chang}, \citenamefont {Kulkarni}, \citenamefont {Lien},\ and\
  \citenamefont {Watson}}]{hari1998nmr}%
  \BibitemOpen
  \bibfield  {author} {\bibinfo {author} {\bibfnamefont {P.}~\bibnamefont
  {Hari}}, \bibinfo {author} {\bibfnamefont {C.}~\bibnamefont {Chang}},
  \bibinfo {author} {\bibfnamefont {R.}~\bibnamefont {Kulkarni}}, \bibinfo
  {author} {\bibfnamefont {J.}~\bibnamefont {Lien}}, \ and\ \bibinfo {author}
  {\bibfnamefont {A.}~\bibnamefont {Watson}},\ }\bibfield  {title} {\enquote
  {\bibinfo {title} {{NMR} characterization of hydrocarbon gas in porous
  media},}\ }\href
  {http://www.sciencedirect.com/science/article/pii/S0730725X98000575}
  {\bibfield  {journal} {\bibinfo  {journal} {Magn. Reson. Imaging}\ }\textbf
  {\bibinfo {volume} {16}},\ \bibinfo {pages} {545--547} (\bibinfo {year}
  {1998})}\BibitemShut {NoStop}%
\bibitem [{\citenamefont {Johnson~Jr}\ and\ \citenamefont
  {Waugh}(1961)}]{johnson1961nuclear}%
  \BibitemOpen
  \bibfield  {author} {\bibinfo {author} {\bibfnamefont {C.}~\bibnamefont
  {Johnson~Jr}}\ and\ \bibinfo {author} {\bibfnamefont {J.}~\bibnamefont
  {Waugh}},\ }\bibfield  {title} {\enquote {\bibinfo {title} {Nuclear
  relaxation in gases: Mixtures of methane and oxygen},}\ }\href
  {http://scitation.aip.org/content/aip/journal/jcp/35/6/10.1063/1.1732204}
  {\bibfield  {journal} {\bibinfo  {journal} {J. Chem. Phys.}\ }\textbf
  {\bibinfo {volume} {35}},\ \bibinfo {pages} {2020} (\bibinfo {year}
  {1961})}\BibitemShut {NoStop}%
\bibitem [{\citenamefont {Venkataramanan}\ \emph {et~al.}(2002)\citenamefont
  {Venkataramanan}, \citenamefont {Song},\ and\ \citenamefont
  {Hurlimann}}]{venkataramanan2002solving}%
  \BibitemOpen
  \bibfield  {author} {\bibinfo {author} {\bibfnamefont {L.}~\bibnamefont
  {Venkataramanan}}, \bibinfo {author} {\bibfnamefont {Y.-Q.}\ \bibnamefont
  {Song}}, \ and\ \bibinfo {author} {\bibfnamefont {M.~D.}\ \bibnamefont
  {Hurlimann}},\ }\bibfield  {title} {\enquote {\bibinfo {title} {Solving
  fredholm integrals of the first kind with tensor product structure in 2 and
  2.5 dimensions},}\ }\href
  {http://ieeexplore.ieee.org/xpl/articleDetails.jsp?arnumber=995059&filter%3DAND%28p_IS_Number%3A21461%29}
  {\bibfield  {journal} {\bibinfo  {journal} {IEEE Trans. Signal Process.}\
  }\textbf {\bibinfo {volume} {50}},\ \bibinfo {pages} {1017--1026} (\bibinfo
  {year} {2002})}\BibitemShut {NoStop}%
\bibitem [{\citenamefont {Hubbard}(1963)}]{hubbard1963theory}%
  \BibitemOpen
  \bibfield  {author} {\bibinfo {author} {\bibfnamefont {P.~S.}\ \bibnamefont
  {Hubbard}},\ }\bibfield  {title} {\enquote {\bibinfo {title} {Theory of
  nuclear magnetic relaxation by spin-rotational interactions in liquids},}\
  }\href {http://journals.aps.org/pr/abstract/10.1103/PhysRev.131.1155}
  {\bibfield  {journal} {\bibinfo  {journal} {Phys. Rev.}\ }\textbf {\bibinfo
  {volume} {131}},\ \bibinfo {pages} {1155} (\bibinfo {year}
  {1963})}\BibitemShut {NoStop}%
\bibitem [{\citenamefont {Abragam}(1961)}]{abragam1961principles}%
  \BibitemOpen
  \bibfield  {author} {\bibinfo {author} {\bibfnamefont {A.}~\bibnamefont
  {Abragam}},\ }\href@noop {} {\emph {\bibinfo {title} {The principles of
  nuclear magnetism}}},\ \bibinfo {number} {32}\ (\bibinfo  {publisher} {Oxford
  university press},\ \bibinfo {year} {1961})\BibitemShut {NoStop}%
\bibitem [{\citenamefont {Oosting}\ and\ \citenamefont
  {Trappeniers}(1971)}]{oosting1971proton}%
  \BibitemOpen
  \bibfield  {author} {\bibinfo {author} {\bibfnamefont {P.~H.}\ \bibnamefont
  {Oosting}}\ and\ \bibinfo {author} {\bibfnamefont {N.}~\bibnamefont
  {Trappeniers}},\ }\bibfield  {title} {\enquote {\bibinfo {title}
  {Proton-spin-lattice relaxation and self-diffusion in methanes: {IV}.
  self-diffusion in methane},}\ }\href
  {http://www.sciencedirect.com/science/article/pii/0031891471900504}
  {\bibfield  {journal} {\bibinfo  {journal} {Physica}\ }\textbf {\bibinfo
  {volume} {51}},\ \bibinfo {pages} {418--431} (\bibinfo {year}
  {1971})}\BibitemShut {NoStop}%
\bibitem [{\citenamefont {Gerritsma}\ \emph {et~al.}(1971)\citenamefont
  {Gerritsma}, \citenamefont {Oosting},\ and\ \citenamefont
  {Trappeniers}}]{gerritsma1971proton}%
  \BibitemOpen
  \bibfield  {author} {\bibinfo {author} {\bibfnamefont {C.}~\bibnamefont
  {Gerritsma}}, \bibinfo {author} {\bibfnamefont {P.}~\bibnamefont {Oosting}},
  \ and\ \bibinfo {author} {\bibfnamefont {N.}~\bibnamefont {Trappeniers}},\
  }\bibfield  {title} {\enquote {\bibinfo {title} {Proton-spin-lattice
  relaxation and self-diffusion in methanes: {II}. experimental results for
  proton-spin-lattice relaxation times},}\ }\href
  {http://www.sciencedirect.com/science/article/pii/0031891471900486}
  {\bibfield  {journal} {\bibinfo  {journal} {Physica}\ }\textbf {\bibinfo
  {volume} {51}},\ \bibinfo {pages} {381--394} (\bibinfo {year}
  {1971})}\BibitemShut {NoStop}%
\bibitem [{\citenamefont {Song}\ \emph {et~al.}(2002)\citenamefont {Song},
  \citenamefont {Venkataramanan}, \citenamefont {H{\"u}rlimann}, \citenamefont
  {Flaum}, \citenamefont {Frulla},\ and\ \citenamefont {Straley}}]{song2002t}%
  \BibitemOpen
  \bibfield  {author} {\bibinfo {author} {\bibfnamefont {Y.-Q.}\ \bibnamefont
  {Song}}, \bibinfo {author} {\bibfnamefont {L.}~\bibnamefont
  {Venkataramanan}}, \bibinfo {author} {\bibfnamefont {M.}~\bibnamefont
  {H{\"u}rlimann}}, \bibinfo {author} {\bibfnamefont {M.}~\bibnamefont
  {Flaum}}, \bibinfo {author} {\bibfnamefont {P.}~\bibnamefont {Frulla}}, \
  and\ \bibinfo {author} {\bibfnamefont {C.}~\bibnamefont {Straley}},\
  }\bibfield  {title} {\enquote {\bibinfo {title} {${T}_1$-${T}_2$ correlation
  spectra obtained using a fast two-dimensional laplace inversion},}\ }\href
  {http://www.sciencedirect.com/science/article/pii/S1090780701924747}
  {\bibfield  {journal} {\bibinfo  {journal} {J. Magn. Reson.}\ }\textbf
  {\bibinfo {volume} {154}},\ \bibinfo {pages} {261--268} (\bibinfo {year}
  {2002})}\BibitemShut {NoStop}%
\bibitem [{\citenamefont {Sho-Wei}\ \emph {et~al.}(2000)\citenamefont
  {Sho-Wei}, \citenamefont {Hirasaki}, \citenamefont {House},\ and\
  \citenamefont {Riki}}]{sho2000correlations}%
  \BibitemOpen
  \bibfield  {author} {\bibinfo {author} {\bibfnamefont {L.}~\bibnamefont
  {Sho-Wei}}, \bibinfo {author} {\bibfnamefont {G.}~\bibnamefont {Hirasaki}},
  \bibinfo {author} {\bibfnamefont {W.}~\bibnamefont {House}}, \ and\ \bibinfo
  {author} {\bibfnamefont {K.}~\bibnamefont {Riki}},\ }\bibfield  {title}
  {\enquote {\bibinfo {title} {Correlations of {NMR} relaxation time with
  viscosity, diffusivity, and gas/oil ratio of methane/hydrocarbon mixtures},}\
  }in\ \href {https://www.onepetro.org/conference-paper/SPE-63217-MS} {\emph
  {\bibinfo {booktitle} {Proceedings of the SPE Annual Technical Conference and
  Exhibition}}}\ (\bibinfo {year} {2000})\BibitemShut {NoStop}%
\bibitem [{\citenamefont {Bevington}\ and\ \citenamefont
  {Robinson}(1969)}]{bevington1969data}%
  \BibitemOpen
  \bibfield  {author} {\bibinfo {author} {\bibfnamefont {P.~R.}\ \bibnamefont
  {Bevington}}\ and\ \bibinfo {author} {\bibfnamefont {D.~K.}\ \bibnamefont
  {Robinson}},\ }\href@noop {} {\emph {\bibinfo {title} {Data reduction and
  error analysis for the physical sciences}}},\ Vol.\ \bibinfo {volume} {336}\
  (\bibinfo  {publisher} {McGraw-Hill New York},\ \bibinfo {year}
  {1969})\BibitemShut {NoStop}%
\bibitem [{\citenamefont {Krynicki}\ \emph {et~al.}(1977)\citenamefont
  {Krynicki}, \citenamefont {Powles},\ and\ \citenamefont
  {Rigamonti}}]{krynicki1977proton}%
  \BibitemOpen
  \bibfield  {author} {\bibinfo {author} {\bibfnamefont {K.}~\bibnamefont
  {Krynicki}}, \bibinfo {author} {\bibfnamefont {J.}~\bibnamefont {Powles}}, \
  and\ \bibinfo {author} {\bibfnamefont {A.}~\bibnamefont {Rigamonti}},\
  }\bibfield  {title} {\enquote {\bibinfo {title} {Proton spin relaxation due
  to critical fluctuations in liquid chloroform},}\ }\href
  {http://www.tandfonline.com/doi/abs/10.1080/00268977700102911} {\bibfield
  {journal} {\bibinfo  {journal} {Mol. Phys.}\ }\textbf {\bibinfo {volume}
  {34}},\ \bibinfo {pages} {1717--1726} (\bibinfo {year} {1977})}\BibitemShut
  {NoStop}%
\bibitem [{\citenamefont {Topol}\ \emph {et~al.}(1979)\citenamefont {Topol},
  \citenamefont {Krynicki},\ and\ \citenamefont {Powles}}]{topol1979nuclear}%
  \BibitemOpen
  \bibfield  {author} {\bibinfo {author} {\bibfnamefont {R.}~\bibnamefont
  {Topol}}, \bibinfo {author} {\bibfnamefont {K.}~\bibnamefont {Krynicki}}, \
  and\ \bibinfo {author} {\bibfnamefont {J.}~\bibnamefont {Powles}},\
  }\bibfield  {title} {\enquote {\bibinfo {title} {Nuclear spin relaxation in
  molecular fluids near the critical point},}\ }\href
  {http://www.tandfonline.com/doi/abs/10.1080/00268977900101211?journalCode=tmph20}
  {\bibfield  {journal} {\bibinfo  {journal} {Mol. Phys.}\ }\textbf {\bibinfo
  {volume} {37}},\ \bibinfo {pages} {1655--1659} (\bibinfo {year}
  {1979})}\BibitemShut {NoStop}%
\bibitem [{\citenamefont {Resing}(1965)}]{resing1965apparent}%
  \BibitemOpen
  \bibfield  {author} {\bibinfo {author} {\bibfnamefont {H.}~\bibnamefont
  {Resing}},\ }\bibfield  {title} {\enquote {\bibinfo {title} {Apparent
  phase-transition effect in the {NMR} spin-spin relaxation time caused by a
  distribution of correlation times},}\ }\href
  {http://scitation.aip.org/content/aip/journal/jcp/43/2/10.1063/1.1696791}
  {\bibfield  {journal} {\bibinfo  {journal} {J. Chem. Phys.}\ }\textbf
  {\bibinfo {volume} {43}},\ \bibinfo {pages} {669--678} (\bibinfo {year}
  {1965})}\BibitemShut {NoStop}%
\bibitem [{\citenamefont {Kimmich}\ and\ \citenamefont
  {Weber}(1993)}]{kimmich1993nmr}%
  \BibitemOpen
  \bibfield  {author} {\bibinfo {author} {\bibfnamefont {R.}~\bibnamefont
  {Kimmich}}\ and\ \bibinfo {author} {\bibfnamefont {H.~W.}\ \bibnamefont
  {Weber}},\ }\bibfield  {title} {\enquote {\bibinfo {title} {{NMR} relaxation
  and the orientational structure factor},}\ }\href {\doibase
  10.1103/PhysRevB.47.11788} {\bibfield  {journal} {\bibinfo  {journal} {Phys.
  Rev. B}\ }\textbf {\bibinfo {volume} {47}},\ \bibinfo {pages} {11788--11794}
  (\bibinfo {year} {1993})}\BibitemShut {NoStop}%
\bibitem [{\citenamefont {Levitz}\ \emph {et~al.}(2000)\citenamefont {Levitz},
  \citenamefont {Korb}, \citenamefont {Quynh},\ and\ \citenamefont
  {Bryant}}]{levitz2000probing}%
  \BibitemOpen
  \bibfield  {author} {\bibinfo {author} {\bibfnamefont {P.}~\bibnamefont
  {Levitz}}, \bibinfo {author} {\bibfnamefont {J.}~\bibnamefont {Korb}},
  \bibinfo {author} {\bibfnamefont {A.~V.}\ \bibnamefont {Quynh}}, \ and\
  \bibinfo {author} {\bibfnamefont {R.}~\bibnamefont {Bryant}},\ }\bibfield
  {title} {\enquote {\bibinfo {title} {Probing dynamics of water molecules in
  mesoscopic disordered media by {NMR} dispersion and 3d simulations in
  reconstructed confined geometries},}\ }in\ \href
  {http://journals.cambridge.org/action/displayAbstract?fromPage=online&aid=8225415&fileId=S1946427400622200}
  {\emph {\bibinfo {booktitle} {MRS Proceedings}}}\ (\bibinfo {organization}
  {Cambridge Univ Press},\ \bibinfo {year} {2000})\BibitemShut {NoStop}%
\bibitem [{\citenamefont {Callaghan}(2011)}]{callaghan2011translational}%
  \BibitemOpen
  \bibfield  {author} {\bibinfo {author} {\bibfnamefont {P.~T.}\ \bibnamefont
  {Callaghan}},\ }\href@noop {} {\emph {\bibinfo {title} {Translational
  Dynamics and Magnetic Resonance: Principles of Pulsed Gradient Spin Echo
  {NMR}}}}\ (\bibinfo  {publisher} {Oxford University Press},\ \bibinfo {year}
  {2011})\BibitemShut {NoStop}%
\bibitem [{\citenamefont {Dozier}\ \emph {et~al.}(1986)\citenamefont {Dozier},
  \citenamefont {Drake},\ and\ \citenamefont {Klafter}}]{dozier1986self}%
  \BibitemOpen
  \bibfield  {author} {\bibinfo {author} {\bibfnamefont {W.~D.}\ \bibnamefont
  {Dozier}}, \bibinfo {author} {\bibfnamefont {J.~M.}\ \bibnamefont {Drake}}, \
  and\ \bibinfo {author} {\bibfnamefont {J.}~\bibnamefont {Klafter}},\
  }\bibfield  {title} {\enquote {\bibinfo {title} {Self-diffusion of a molecule
  in porous vycor glass},}\ }\href {\doibase 10.1103/PhysRevLett.56.197}
  {\bibfield  {journal} {\bibinfo  {journal} {Phys. Rev. Lett.}\ }\textbf
  {\bibinfo {volume} {56}},\ \bibinfo {pages} {197--200} (\bibinfo {year}
  {1986})}\BibitemShut {NoStop}%
\bibitem [{\citenamefont {Van~Landeghem}\ \emph {et~al.}(2010)\citenamefont
  {Van~Landeghem}, \citenamefont {Haber}, \citenamefont {D'espinose
  De~Lacaillerie},\ and\ \citenamefont {Bl{\"u}mich}}]{van2010analysis}%
  \BibitemOpen
  \bibfield  {author} {\bibinfo {author} {\bibfnamefont {M.}~\bibnamefont
  {Van~Landeghem}}, \bibinfo {author} {\bibfnamefont {A.}~\bibnamefont
  {Haber}}, \bibinfo {author} {\bibfnamefont {J.-B.}\ \bibnamefont {D'espinose
  De~Lacaillerie}}, \ and\ \bibinfo {author} {\bibfnamefont {B.}~\bibnamefont
  {Bl{\"u}mich}},\ }\bibfield  {title} {\enquote {\bibinfo {title} {Analysis of
  multisite {2D} relaxation exchange {NMR}},}\ }\href
  {http://onlinelibrary.wiley.com/doi/10.1002/cmr.a.20157/full} {\bibfield
  {journal} {\bibinfo  {journal} {Concepts Magn. Reson., Part A}\ }\textbf
  {\bibinfo {volume} {36}},\ \bibinfo {pages} {153--169} (\bibinfo {year}
  {2010})}\BibitemShut {NoStop}%
\bibitem [{\citenamefont {Lee}\ \emph {et~al.}(1993)\citenamefont {Lee},
  \citenamefont {Labadie}, \citenamefont {Springer~Jr},\ and\ \citenamefont
  {Harbison}}]{lee1993two}%
  \BibitemOpen
  \bibfield  {author} {\bibinfo {author} {\bibfnamefont {J.~H.}\ \bibnamefont
  {Lee}}, \bibinfo {author} {\bibfnamefont {C.}~\bibnamefont {Labadie}},
  \bibinfo {author} {\bibfnamefont {C.~S.}\ \bibnamefont {Springer~Jr}}, \ and\
  \bibinfo {author} {\bibfnamefont {G.~S.}\ \bibnamefont {Harbison}},\
  }\bibfield  {title} {\enquote {\bibinfo {title} {Two-dimensional inverse
  laplace transform {NMR}: altered relaxation times allow detection of exchange
  correlation},}\ }\href {http://pubs.acs.org/doi/abs/10.1021/ja00070a022}
  {\bibfield  {journal} {\bibinfo  {journal} {J. Am. Chem. Soc.}\ }\textbf
  {\bibinfo {volume} {115}},\ \bibinfo {pages} {7761--7764} (\bibinfo {year}
  {1993})}\BibitemShut {NoStop}%
\bibitem [{\citenamefont {Washburn}\ and\ \citenamefont
  {Callaghan}(2006)}]{washburn2006tracking}%
  \BibitemOpen
  \bibfield  {author} {\bibinfo {author} {\bibfnamefont {K.~E.}\ \bibnamefont
  {Washburn}}\ and\ \bibinfo {author} {\bibfnamefont {P.~T.}\ \bibnamefont
  {Callaghan}},\ }\bibfield  {title} {\enquote {\bibinfo {title} {Tracking pore
  to pore exchange using relaxation exchange spectroscopy},}\ }\href {\doibase
  10.1103/PhysRevLett.97.175502} {\bibfield  {journal} {\bibinfo  {journal}
  {Phys. Rev. Lett.}\ }\textbf {\bibinfo {volume} {97}},\ \bibinfo {pages}
  {175502} (\bibinfo {year} {2006})}\BibitemShut {NoStop}%
\bibitem [{\citenamefont {Kleinberg}\ and\ \citenamefont
  {Vinegar}(1996)}]{kleinberg1996nmr}%
  \BibitemOpen
  \bibfield  {author} {\bibinfo {author} {\bibfnamefont {R.}~\bibnamefont
  {Kleinberg}}\ and\ \bibinfo {author} {\bibfnamefont {H.}~\bibnamefont
  {Vinegar}},\ }\bibfield  {title} {\enquote {\bibinfo {title} {{NMR}
  properties of reservoir fluids},}\ }\href
  {https://www.onepetro.org/journal-paper/SPWLA-1996-v37n6a4} {\bibfield
  {journal} {\bibinfo  {journal} {Log Analyst}\ }\textbf {\bibinfo {volume}
  {37}},\ \bibinfo {pages} {20--32} (\bibinfo {year} {1996})}\BibitemShut
  {NoStop}%
\bibitem [{\citenamefont {Gensterblum}\ \emph {et~al.}(2009)\citenamefont
  {Gensterblum}, \citenamefont {Van~Hemert}, \citenamefont {Billemont},
  \citenamefont {Busch}, \citenamefont {Charri{\'e}re}, \citenamefont {Li},
  \citenamefont {Krooss}, \citenamefont {De~Weireld}, \citenamefont {Prinz},\
  and\ \citenamefont {Wolf}}]{gensterblum2009european}%
  \BibitemOpen
  \bibfield  {author} {\bibinfo {author} {\bibfnamefont {Y.}~\bibnamefont
  {Gensterblum}}, \bibinfo {author} {\bibfnamefont {P.}~\bibnamefont
  {Van~Hemert}}, \bibinfo {author} {\bibfnamefont {P.}~\bibnamefont
  {Billemont}}, \bibinfo {author} {\bibfnamefont {A.}~\bibnamefont {Busch}},
  \bibinfo {author} {\bibfnamefont {D.}~\bibnamefont {Charri{\'e}re}}, \bibinfo
  {author} {\bibfnamefont {D.}~\bibnamefont {Li}}, \bibinfo {author}
  {\bibfnamefont {B.}~\bibnamefont {Krooss}}, \bibinfo {author} {\bibfnamefont
  {G.}~\bibnamefont {De~Weireld}}, \bibinfo {author} {\bibfnamefont
  {D.}~\bibnamefont {Prinz}}, \ and\ \bibinfo {author} {\bibfnamefont {K.-H.}\
  \bibnamefont {Wolf}},\ }\bibfield  {title} {\enquote {\bibinfo {title}
  {European inter-laboratory comparison of high pressure ${CO}_2$ sorption
  isotherms. {I}: Activated carbon},}\ }\href
  {http://www.sciencedirect.com/science/article/pii/S0008622309003935}
  {\bibfield  {journal} {\bibinfo  {journal} {Carbon}\ }\textbf {\bibinfo
  {volume} {47}},\ \bibinfo {pages} {2958--2969} (\bibinfo {year}
  {2009})}\BibitemShut {NoStop}%
\bibitem [{\citenamefont {Langmuir}(1918)}]{langmuir1918adsorption}%
  \BibitemOpen
  \bibfield  {author} {\bibinfo {author} {\bibfnamefont {I.}~\bibnamefont
  {Langmuir}},\ }\bibfield  {title} {\enquote {\bibinfo {title} {The adsorption
  of gases on plane surfaces of glass, mica and platinum.}}\ }\href
  {http://pubs.acs.org/doi/abs/10.1021/ja02242a004} {\bibfield  {journal}
  {\bibinfo  {journal} {J. Am. Chem. Soc.}\ }\textbf {\bibinfo {volume} {40}},\
  \bibinfo {pages} {1361--1403} (\bibinfo {year} {1918})}\BibitemShut {NoStop}%
\bibitem [{\citenamefont {Brunauer}\ \emph {et~al.}(1938)\citenamefont
  {Brunauer}, \citenamefont {Emmett},\ and\ \citenamefont
  {Teller}}]{brunauer1938adsorption}%
  \BibitemOpen
  \bibfield  {author} {\bibinfo {author} {\bibfnamefont {S.}~\bibnamefont
  {Brunauer}}, \bibinfo {author} {\bibfnamefont {P.~H.}\ \bibnamefont
  {Emmett}}, \ and\ \bibinfo {author} {\bibfnamefont {E.}~\bibnamefont
  {Teller}},\ }\bibfield  {title} {\enquote {\bibinfo {title} {Adsorption of
  gases in multimolecular layers},}\ }\href
  {http://pubs.acs.org/doi/abs/10.1021/ja01269a023} {\bibfield  {journal}
  {\bibinfo  {journal} {J. Am. Chem. Soc.}\ }\textbf {\bibinfo {volume} {60}},\
  \bibinfo {pages} {309--319} (\bibinfo {year} {1938})}\BibitemShut {NoStop}%
\bibitem [{\citenamefont {Press}(2007)}]{press2007numerical}%
  \BibitemOpen
  \bibfield  {author} {\bibinfo {author} {\bibfnamefont {W.~H.}\ \bibnamefont
  {Press}},\ }\href@noop {} {\emph {\bibinfo {title} {Numerical recipes 3rd
  edition: The art of scientific computing}}}\ (\bibinfo  {publisher}
  {Cambridge university press},\ \bibinfo {year} {2007})\BibitemShut {NoStop}%
\bibitem [{\citenamefont {Himeno}\ \emph {et~al.}(2005)\citenamefont {Himeno},
  \citenamefont {Komatsu},\ and\ \citenamefont {Fujita}}]{himeno2005high}%
  \BibitemOpen
  \bibfield  {author} {\bibinfo {author} {\bibfnamefont {S.}~\bibnamefont
  {Himeno}}, \bibinfo {author} {\bibfnamefont {T.}~\bibnamefont {Komatsu}}, \
  and\ \bibinfo {author} {\bibfnamefont {S.}~\bibnamefont {Fujita}},\
  }\bibfield  {title} {\enquote {\bibinfo {title} {High-pressure adsorption
  equilibria of methane and carbon dioxide on several activated carbons},}\
  }\href {http://pubs.acs.org/doi/abs/10.1021/je049786x} {\bibfield  {journal}
  {\bibinfo  {journal} {J. Chem. Eng. Data}\ }\textbf {\bibinfo {volume}
  {50}},\ \bibinfo {pages} {369--376} (\bibinfo {year} {2005})}\BibitemShut
  {NoStop}%
\end{thebibliography}%

\end{document}